\pdfoutput=1
\documentclass[aps,physrev,preprint,groupedaddress,amsmath,amssymb]{revtex4-2}

\usepackage{graphicx}
\usepackage{bm}
\usepackage{float}     % [H] figure placement (used in Results)
\usepackage{caption}   % \captionsetup (used in Appendix A)
\usepackage{placeins}  % \FloatBarrier (used in Results)
\usepackage[utf8]{inputenc}
\usepackage[colorlinks=true,linkcolor=blue,citecolor=blue,urlcolor=blue]{hyperref}
\usepackage[numbered]{bookmark}  % number the PDF outline entries (I., A., ...) to match the body

\newcommand{\ucd}[1]{\overset{\triangledown}{#1}}

\newcommand{\NN}{\mathrm{NN}}

\renewcommand{\TH}{\mathrm{TH}}
\newcommand{\Wi}{\mathrm{Wi}}

\begin{document}

% Optional preprint / report number (top-right of title page in preprint mode):
% \preprint{arXiv:XXXX.XXXXX}   % <-- set your arXiv / institutional preprint number

\title{A GENERIC-Informed Neural Network for Generative Rheology}

\author{Imanol García de Beristain}
\email{imanol.garciad@ehu.eus}
\affiliation{Applied Mathematics Department, Engineering School of Bilbao, University of the Basque Country (UPV/EHU), Plaza Ingeniero Torres Quevedo 1, 48013 Bilbao, Spain}

\author{David Nieto Simavilla}
\affiliation{Departamento de Energía y Combustibles, Escuela Técnica Superior de Ingenieros de Minas y Energía, Universidad Politécnica de Madrid, Calle Alenza 4, 28003 Madrid, Spain}

\author{Marco Ellero}
\affiliation{Basque Center for Applied Mathematics (BCAM), Alameda Mazarredo 14, 48009 Bilbao, Spain}
\affiliation{IKERBASQUE, Basque Foundation for Science, Calle María Díaz de Haro 3, 48013 Bilbao, Spain}

\begin{abstract}
We present GINNGER (GENERIC-Informed Neural Network for GEnerative Rheology), a framework for discovering thermodynamically admissible constitutive equations from microstructural flow data. Unlike existing approaches that learn constitutive operators, GINNGER reformulates constitutive discovery as the learning of the underlying thermodynamic generators from which constitutive equations emerge. Embedding the GENERIC formalism into the learning process restricts the search to constitutive models that satisfy the first and second laws of thermodynamics by construction, eliminating the need to prescribe phenomenological constitutive equations a priori.
A second conceptual advance is a spectral formulation of constitutive discovery in the intrinsic eigenspace of the polymer conformation tensor. This provides a kinematics-independent thermodynamic state space in which the thermodynamic generators depend only on the instantaneous microstructural state, while flow kinematics acts solely as an external driver transporting the system through this space. Consequently, data from shear, extensional, and mixed flows contribute to the discovery of a single constitutive model that generalizes across deformation protocols and flow geometries.

By learning thermodynamic generators rather than high-dimensional constitutive operators, GINNGER substantially reduces the complexity of the discovery problem while preserving thermodynamic consistency, material objectivity, and physical interpretability by construction. The framework is validated using high-fidelity simulations, accurately reconstructing constitutive behavior and predicting complex flow fields across previously unseen geometries. GINNGER establishes a new paradigm for constitutive discovery by combining generative, spectral, and thermodynamically grounded learning within a unified framework for data-driven modeling of complex fluids.

\end{abstract}

\maketitle

\section{Introduction}\label{sec:intro}
Complex fluids such as polymer solutions and melts exhibit a rich flow behavior that strongly 
departs from the Newtonian paradigm. Predicting these flows requires constitutive equations 
that connect macroscopic stresses to the underlying microstructure, in addition to the 
universal balance equations for mass and momentum. For polymeric fluids, such constitutive 
models are often formulated in terms of an evolution equation for a conformation tensor and a closure relation for the non-Newtonian extra stress. From a thermodynamic perspective, the 
General Equation for Non-Equilibrium Reversible–Irreversible Coupling (\textsc{generic}) 
provides a unifying framework to construct such models in a way that is consistent with energy conservation and entropy production. \textsc{generic} has been successfully applied to derive rheological equations for dilute polymer solutions and more complex materials~\cite{Ottinger_book,Grmela_1997, Ilg2009, Ilg2011, Schieber2021}.

Traditional rheological modeling proceeds by postulating a specific functional form for the 
constitutive equation that is grounded in physical intuition and mesoscopic modeling. Then a small set of parameters is calibrated using simple rheometric experiments. Classical texts have established a rich catalogue of such constitutive equations for complex fluids, but 
the resulting models are necessarily a compromise between physical fidelity, tractability, and the limited
range of flows explored experimentally~\cite{Bird1987,Larson1988,Dyson2004}. In 
particular, the calibration is typically performed using steady shear or extensional flows, 
which probe only restricted regions of the conformation and stress space. Extrapolating these 
models to complex three-dimensional flows, transient regimes, or flow configurations very 
different from those used for calibration remains a persistent challenge. Moreover, as the 
number of model parameters grows to capture richer physics (e.g. multiple relaxation times, 
anisotropic mobility), the calibration process becomes increasingly under-determined and 
sensitive to experimental and numerical noise.

Over the past decade, advances in machine learning (ML) and scientific computing have motivated a complementary, data-driven route to constitutive modeling. Rather than postulating a fixed functional form \emph{a priori}, one seeks to infer it from data, typically by training expressive function approximators such as neural networks or Gaussian processes \cite{Seryo2020,zhao2021physics,raissi2017machine}. In the context of rheology and complex fluids, this has led to several distinct lines of research. Some approaches use high-fidelity microscopic or mesoscopic simulations to generate data and then employ regression or Gaussian process models to learn stress–strain-rate relationships under simple flows~\cite{zhao2018,zhao2021,Seryo2020}. Others use sparse regression or operator-learning techniques to identify reduced-order models or closure terms for complex 
suspensions and non-Newtonian fluids~\cite{discovering_brunton_2016,datadriven_rudy_2016}. More recently, physics-informed neural networks (PINNs) have been developed as a flexible tool for 
learning governing equations and constitutive relations directly from data while enforcing the 
underlying partial differential equations as soft constraints during training~\cite{raissi2019physics,Karniadakis2021,Cuomo2022,Cai2021}.

In the rheology community, PINN-based methods and related rheology-informed networks have been explored to (i) learn constitutive parameters of pre-specified models, (ii) reconstruct flow fields from partial measurements, and (iii) represent unknown components of the constitutive law in a non-parametric way~\cite{Mahmou2022,Saadat2022,Mahmou2022b,Young2023}. These approaches 
illustrate the potential of ML and PINNs to combine data with physical structure and to exploit 
diverse sources of information, from rheometric experiments to complex flow simulations \cite{Sankaran2026}. At the same time, a broader body of work in scientific machine learning for complex fluids has emerged. Lennon \emph{et al.} introduced a framework named Rheological Universal 
Differential Equations (RUDEs), which embeds universal function approximators into objective tensorial constitutive equations and uses adjoint-based training to learn extra-stress 
contributions that can be transferred to CFD simulations~\cite{scientific_lennon_2023}. 
Thermodynamics-aware and metriplectic formulations have been proposed to ensure compatibility with non-equilibrium thermodynamics when learning reduced-order or graph-based models, explicitly separating reversible and irreversible dynamics and treating the entropy as a latent learned quantity~\cite{deep_hernndez_2021,thermodynamicsinformed_hernndez_2022,machine_lee_2021}. 
{Closely related to this line of work, recent developments such as the GIMLET framework explicitly embed thermodynamic structure into machine-learning models, enabling generalizable and interpretable learning of constitutive relations while preserving consistency with energy and entropy principles \cite{shiratori2025gimlet}.}
Differentiable non-Newtonian solvers have also been developed, in which tensor-based neural 
networks are trained end-to-end through the CFD solver using velocity measurements, and the 
resulting models serve as digital rheometers from which classical constitutive equations such as Giesekus or Carreau–Yasuda can be identified via Bayesian model 
selection when full field information is not available~\cite{Sunol2026}. Furthermore, benchmarking studies of 
scientific ML operators for flow prediction around complex geometries have assessed neural 
operators, DeepONets, and scalable transformer-based operators, highlighting the role of 
geometry representation (e.g. signed distance functions versus binary masks) and the challenges of out-of-distribution generalization~\cite{Balu2025}. 
{In parallel, recent experimental–data-driven approaches have extracted rheological properties from basic measurements, such as video-based high-throughput viscometry, further expanding the range of accessible training data and tools for ML-based constitutive modelling \cite{arretche2025}.}

Despite this rapid progress, many ML constitutive models still rely on a fixed, hand-chosen 
functional form (for example, a given viscometric function or a specific model like FENE-P or Giesekus) and restrict learning to a small set of parameters. 

In some cases, a finite family of models is used and during the parameter identification process the model that most closely captures the data is chosen. In both cases, these approaches rely on specific (single or multiple) constitutive equations that are pre-defined and therefore, they cannot be classified as fully model-independent. 
{For instance, sparse identification techniques have been successfully used to discover governing equations from data, but they typically rely on a predefined library of candidate terms, which constrains the space of admissible constitutive models \cite{sato2026}.}
Others treat the constitutive mapping as a purely black-box surrogate between kinematic quantities (e.g. local velocity gradient) and stress, paying limited attention to thermodynamic consistency or microstructural interpretations. As a result, these models may exhibit nonphysical behaviour when used to provide predictions outside the training conditions and/or geometry. 

More fundamentally, existing data-driven approaches formulate constitutive discovery as the identification of constitutive operators relating kinematic quantities to stresses or conformation dynamics. Whether these operators are represented through predefined constitutive equations, sparse libraries, tensor-basis expansions, or neural networks, the constitutive equation itself remains the primary object of learning. In the present work, we adopt a fundamentally different perspective. Rather than learning constitutive equations directly, we learn the underlying thermodynamic generators from which thermodynamically admissible constitutive equations emerge. By embedding the GENERIC structure into the learning process, thermodynamic consistency is no longer imposed \emph{a posteriori} but instead defines the admissible space itself.

In recent work, we proposed a \textsc{generic}-guided approach in which a PINN learns the entropy contribution governing the extra stress in dilute polymer solutions~\cite{Nieto2025}. Instead of modeling flow fields, the network approximates the specific polymeric entropy as a function of the conformation tensor, while training is constrained by the \textsc{generic}-compliant evolution equation to ensure thermodynamic consistency. Two strategies were explored: training on analytical {data from} Oldroyd-B solutions under steady rheometric flows, and on in-silico data from CFD simulations of flow around a cylinder. {In these blind tests, the ML framework was tasked with recovering an unknown constitutive law from data, which in this case were generated from the Oldroyd–B model but not revealed to the ML algorithm.} The learned entropy model was validated using a finite-volume solver (RheoTool) \cite{Nieto2025}. The versatility of the proposed framework was further demonstrated by combining it with a Gaussian Process Regression active-learning strategy for data-driven Smooth Particle Hydrodynamics simulations of viscoelastic flows \cite{Dong2026}.

A key feature of that formulation was the realization that constitutive learning is most naturally performed in the spectral representation of the conformation tensor rather than in physical space. In this representation, the polymeric microstructure defines an intrinsic thermodynamic state space that is independent of the particular deformation protocol. Different flow kinematics merely transport the microstructure through this state space, while the thermodynamic generators remain functions only of the instantaneous microstructural state. Consequently, data originating from shear, extensional, and mixed flows can all contribute to the identification of the same constitutive model. 
Building on this observation, the present work adopts the intrinsic spectral state space of the conformation tensor as the natural domain for constitutive discovery. The validity of the learned constitutive model is therefore determined by the region of thermodynamic state space explored during training rather than by the particular flow geometry, deformation protocol, or Weissenberg number used to generate the data.

The framework in \cite{Nieto2025}, however, was limited to {viscoelastic} dynamics with a prescribed linear friction matrix and fixed  relaxation mechanism through a single constant time scale. While the network learned the entropy, the friction matrix was assumed known and linear in the conformation tensor, as derived for dilute suspensions~\cite{vazquez2009b, Ottinger_book}. This restricts applicability to more realistic systems, where nonlinear and anisotropic mobility is required to capture hydrodynamic interactions, shear-thinning, and/or elastic instabilities. Moreover, prescribing the form of the friction matrix requires \emph{a priori} specification of relaxation parameters, which are often difficult to infer uniquely from data.

To the best of our knowledge, existing ML approaches to rheological modelling have not been used to automatically discover general tensorial constitutive equations within a thermodynamically consistent framework.  Accordingly, the present framework is built upon two complementary conceptual advances. First, constitutive discovery is reformulated as the identification of thermodynamic generators rather than constitutive operators. Second, this discovery is performed in the intrinsic spectral state space of the conformation tensor rather than in flow-dependent physical coordinates. Together, these ideas transform the search for high-dimensional constitutive equations into the identification of a small number of thermodynamic functions defined over a universal microstructural state space.

\begin{figure}[hbt!]
    \centering
    \includegraphics[width=1.0\linewidth]{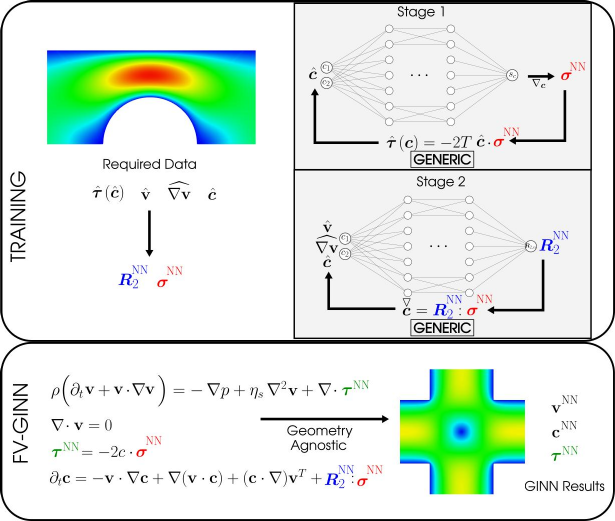}
    \caption{Schematic of GINNGER methodology. Top: Two stage training procedure. Bottom: Rheotool implementation of the learned constitutive model. }
    \label{fig:sketch}
\end{figure}

Here, we propose a \textsc{generic}-guided ML framework for the automatic discovery of such constitutive laws where the training is divided into two stages as depicted in Figure \ref{fig:sketch}. In \textbf{Stage~1}, a neural network is trained on polymeric stress and conformation tensor data {$\bm{c}$} to infer the polymeric configurational entropy {$s_c(\bm{c})$} and its thermodynamic conjugate variable {$\bm{\sigma}(\bm{c})= \partial s_c(\bm{c})/\partial \bm{c}$}  without prescribing relaxation time scales. This extends our previous entropy-learning approach~\cite{Nieto2025} by explicitly recovering $\bm{\sigma}$ from {stress data} $\bm{\tau}$ while enforcing \textsc{generic}'s  {degeneracy condition}. In \textbf{Stage~2}, the friction matrix is treated as an unknown anisotropic second-rank tensorial operator {$\bm{R}_2(\bm{c})$} governing the irreversible dynamics. A second network is trained on conformation dynamics to learn this operator, using the Stage~1 model and enforcing the upper-convected \textsc{generic} equation for the conformation tensor to ensure {evolutionary} thermodynamic consistency. For simplicity, we assume affine deformation, avoiding the need to determine a slip coefficient~\cite{Nieto2023}.

This two-stage procedure yields a \textsc{generic}-guided constitutive model in which neural network representations of the entropy thermodynamic conjugate tensor $\bm{\sigma}^{\rm NN}$ and the anisotropic friction operator $\bm{R}_2^{\rm NN}$ are learned separately but consistently. Stage~1 identifies the thermodynamic driving forces, while Stage~2 identifies the irreversible relaxation mechanism. Together, these neural networks learn the thermodynamic generators defining the constitutive model rather than the constitutive equation itself, eliminating the need for \emph{a priori} specification of relaxation mechanisms or phenomenological model forms. By preserving the \textsc{generic} structure, the resulting framework guarantees compatibility with energy conservation, entropy production, and material objectivity by construction.
A general sketch of how the learned model (composed of several neural networks) is implemented into the RheoTool finite-volume solver is provided in Figure \ref{fig:sketch}.

More broadly, GINNGER introduces a generative and spectral formulation of constitutive discovery. Generative, because thermodynamic generators rather than constitutive operators are learned directly from data; spectral, because these generators are identified in the intrinsic spectral representation of the conformation tensor, where they become independent of the particular flow kinematics. Beyond reducing the dimensionality of the optimization problem, this reformulation provides a physically interpretable framework in which constitutive equations emerge naturally from the learned thermodynamic structure.

\section{GENERIC formulation and numerical setup}
 \label{sec:generic_giesekus}

Assuming that the second-rank conformational tensor $\bm c=\langle \bm q \bm q\rangle$ (second moment of the polymer end-to-end vector distribution) is considered as a complete variable describing the microstructure (physical state system $x$), its most general (\textsc{GENERIC}-compliant) evolution equation reads \cite{Ottinger_book}

\begin{equation}\label{generic_c}
\ucd{\bm c}  = -\frac{{\bm R}_2}{T}  : \frac{\partial f_c}{\partial {\bm c}} 
\end{equation}
where $f_c({\bm c})=\epsilon_c({\bm c}) - Ts_c({\bm c})$ is the configurational Helmholtz free energy density, $T$ is the temperature, and $\epsilon_c( {\bm c})$ and $s_c({\bm c})$ are the configurational energy and entropy densities. ${\bm R}_2({\bf c})$ is a 4th (commonly 2nd) rank tensorial function defining the friction matrix.
The first term in Eq. (\ref{generic_c}) is the upper convective derivative 
\begin{equation}
        \ucd{(..)} 
 = \partial_t {(..)} + {\mathbf{v}} \cdot {\bm \nabla} {(..)} - \bm {\kappa} \cdot {(..)}
- {(..)} \cdot \bm {\kappa}^T    \label{eq:ucd}
\end{equation}
which simply reflects the {\bf affine deformation} of the suspended polymer chains. That is, it describes the kinematic evolution of the conformation tensor under the velocity gradient $\boldsymbol{\kappa}=\nabla \mathbf{v}^T$.

By assuming only entropic contribution, Eq. (\ref{generic_c}) reduces to
\begin{equation}
       \ucd{\bm c} 
 = {\bm R}_2  : \frac{\partial s_c}{\partial {\bm c}}.   \label{eq:generic_c2}
\end{equation}
To summarize, at this point we used three assumptions. First, the evolution of the microstructure can be described by the use of the conformation tensor and a second rank tensor $\bm R_2(\bm c)$. Second, the affine flow kinematics are described by an upper convected derivative (i.e., no slip). And third, the stress and the irreversible terms of the conformation tensor evolution equation depend only on the entropy density and not on configurational energetic terms \cite{Ionescu2008, Nieto2018} (i.e., only entropic models are considered).   

From \textsc{GENERIC}'s degeneracy condition (enforcing that the reversible Poisson matrix does not affect entropy evolution, i.e. it does not produce any net entropy increase), one obtains the {\bf stress-microstructure compatibility equation}. 
 \begin{equation}
 {\bm \tau} ({\bm c})
 = - 2 T  \, {\bm c} \cdot  \frac{\partial s_c}{\partial {\bm c}} \label{eq:generic_stress}
\end{equation}
Note that the stress relation is completely general and does not contain information on the friction matrix ${\bm R}_2$, but only on the irreversible entropic contribution $s_c$.

In order to close the system in Eqs. (\ref{eq:generic_c2}) and (\ref{eq:generic_stress}), we only need expressions for the friction matrix ${\bm R}_2 ( {\bm c} ) $ and the configurational scalar entropy $s_c( {\bm c} )$. Provided that we have analytical expressions for these relations, we can examine different existing models through the thermodynamics-compliant lens of \textsc{GENERIC}. Notice that in more complex hierarchical approaches, the evolution equation at the 2nd rank level $\bm c \propto \langle \bm q \bm q\rangle$ might contain terms involving higher order-tensors (i.e., $\langle \bm q \bm q \bm q \bm q\rangle$, etc) \cite{Pavelka2018}. 
In our approach, we search for the best match that produces a data-driven closure at the single $\bm c$-level in a thermodynamically-consistent way. A few analytical model examples are included in Appendix~\ref{sec:GENERIC_formulation}.

\section{GINNGER approach}

The current work provides data-driven solutions where {the tensorial functions} $\bm \sigma(\bm{c})$ and $\bm R_2(\bm{c})$, {and therefore $\bm{\tau}$}, are modelled with neural networks {($\bm \sigma^\NN$, $\bm R_2^\NN$ and $\bm{\tau}^{\NN}$)}. For validation purposes and without loss of generality, in this work we generate high-fidelity training data using the well-known non-linear anisotropic Giesekus constitutive model. Data generated via this model showcases the applicability of the new framework and it can be easily generalized to many other thermodynamically consistent tensorial models \cite{Schieber2021}. The GINNGER $\bm \sigma^\NN$ and $\bm R_2^\NN$ are trained on these data and their predictions are subsequently validated against the responses of the ground-truth Giesekus constitutive model $\bm \sigma^\TH$ and $\bm R_2^\TH$. For all the constitutive modeling (NN or theoretical), we use the same finite-volume solver and variable-coupling strategy {with the only difference being} that the entropy conjugate and the friction matrix entering the FV discretization are computed analytically for the validation ground-truth model or predicted by the {trained} NN models.

The numerical setup is based on the RheoTool solver, where the incompressible momentum conservation equations, written in the solver’s dimensional units, are discretized via SIMPLEC \cite{PIMENTA2017}. In particular, we work with a fully hydrodynamically coupled scheme where mass/momentum conservation are used
\begin{align}
  \nabla\!\cdot\mathbf{v} &= 0, \label{eq:mass} \\
  \rho\Big(\partial_t\mathbf{v} + \mathbf{v}\!\cdot\!\nabla\mathbf{v}\Big)
  &= -\,\nabla p
     + \eta_s\,\nabla^2\mathbf{v}
     + \nabla\!\cdot \bm\tau, \label{eq:momentum}
\end{align}
where $\mathbf{v}$ is the velocity vector, $p$ is the pressure, $\rho$ is the fluid density, $\eta_s$ is the solvent viscosity, and $\bm \tau$ is the polymeric contribution to the extra-stress tensor, defined by (\ref{eq:generic_stress}) and equipped with (\ref{eq:generic_c2}).

{The FV numerical algorithm employs a segregated SIMPLEC approach. Each time step consists of a velocity-pressure correction with an explicit treatment of the non-Newtonian stress, followed by a constitutive update via the GENERIC framework in Eqs. (\ref{evol_GIESEKUS})--(\ref{eq:stress_GIESEKUS}). The conformation equation is solved using a stabilized stress-velocity coupling scheme. Simulations are run until the conformation tensor converges to a steady-state solution with a residual of $10^{-5}$.}

\subsection{Spectral State-Space Learning}
\label{sec:data_pipeline}

To train the neural networks, we require data for the conformation tensor, polymeric stress, and velocity gradients, which in the present work are generated \emph{in silico} using RheoTool simulations of the ground-truth constitutive model to be discovered (here the Giesekus model). We have used hats to denote $\hat{\bm c}$, $\hat{\bm \tau}$, $\hat{\mathbf v}$ and $\widehat{\nabla \mathbf v}$ as the simulation data and distinguish it from predicted quantities. Nevertheless, the proposed methodology is independent of the origin of the data and can, in principle, be trained using experimental measurements if the corresponding microstructural information is available.

A key conceptual ingredient of GINNGER is that the generative constitutive discovery is performed in the \emph{spectral state space} of the conformation tensor rather than in physical space. Within the \textsc{GENERIC} framework, the thermodynamic generators—the entropy conjugate tensor $\bm{\sigma}$ and the friction tensor $\bm{R}_2$—are tensorial functions of the conformation tensor and therefore diagonalize in the same eigenbasis \cite{Nieto2023,Nieto2025}. The eigenvalues of $\bm c$ thus provide intrinsic coordinates for the thermodynamic state of the polymer microstructure. In this representation, the thermodynamic generators depend only on the instantaneous microstructural state, whereas the flow kinematics acts solely as an external driver transporting the system through this state space. Consequently, shear, extensional and mixed flows correspond to different trajectories through the same spectral thermodynamic state space, allowing heterogeneous datasets to contribute to the identification of a single constitutive model.

An immediate consequence of this reformulation is that the constitutive problem can be expressed in terms of the spectral evolution equations

\begin{equation}
\dot{c}_\alpha  = 2 c_\alpha \kappa_{\alpha\alpha}
 +    R_2^\alpha  \sigma_\alpha ,
\label{eq:eigenvalue_general}
\end{equation}

where $c_\alpha$, $\sigma_\alpha$, and $R^\alpha_2$ denote the eigenvalues of $\bm c=\sum_\alpha c_\alpha {\bf u}_\alpha {\bf u}_\alpha$, $\bm \sigma=\sum_\alpha \sigma_\alpha {\bf u}_\alpha {\bf u}_\alpha$, and $\bm R_2=\sum_\alpha R_{2}^{\alpha} {\bf u}_\alpha {\bf u}_\alpha$, respectively, and $\kappa_{\alpha\alpha}= {\bf u}_\alpha \cdot {\bm \kappa} \cdot{\bf u}_\alpha$ is the transposed velocity gradient ${\bm \kappa}={\bm \nabla \bm v}^T$ projected into the $\bm c$ eigenbasis ${\bf u}_\alpha$.  Equation~(\ref{eq:eigenvalue_general}) remains valid for any entropic constitutive model in which $\bm R_2$ and $\bm \sigma$ are tensorial functions of $\bm c$. Additional details of the spectral decomposition are provided in Appendix~\ref{sec:GENERIC_formulation}. Beyond reducing the dimensionality of the learning problem, this formulation naturally separates the intrinsic thermodynamic description of the material from the external flow kinematics, making the learned constitutive model independent of the particular deformation protocol.

Since the CFD mesh contains many points near equilibrium (\(c_1\approx1\)) and comparatively few highly stretched configurations, the sampling strategy is likewise performed in the spectral state space. A three-step sampling pipeline is employed: (i) Chebyshev-node selection along the data manifold to ensure broad coverage of the thermodynamic states, (ii) farthest-point sampling to avoid clustering, and (iii) enrichment of the upper 1\% of the $c_1$ distribution to improve the representation of highly stretched configurations. This yields a reduced but well-conditioned training dataset that uniformly covers the physically relevant thermodynamic state space rather than particular flow geometries or deformation protocols.

The following sections describe the two-stage learning procedure used to identify the thermodynamic generators $\bm{\sigma}^{\rm NN}$ and $\bm{R}_2^{\rm NN}$ from this spectral state-space representation of the polymer microstructure.

\subsection{Stage 1: learning the entropy conjugate}
\label{subsec:stage1}

We construct a scalar {thermodynamic} potential \(s^{\NN}({\bm c})\) and its conjugate
$\bm \sigma^\NN({\bm c})$ from conformation data \(\hat {\bm c}\) and stress data \(\hat {\bm \tau}\). We use the residual on the {stress-microstructure relation} directly, instead of an {evolutionary} constitutive-residual loss, to avoid relying on a known polymer relaxation time $\lambda$. The stress and the entropy
conjugate are related via the GENERIC relation given by the {stress relation} in Eq. (\ref{eq:generic_stress}) which, as discussed earlier, follows directly from the GENERIC degeneracy condition expressed as $L \cdot \frac{\partial S}{\partial x} = 0$  \cite{Ottinger_book}.
{Remarkably}, this condition does not {rely on the specific} structure of the friction matrix $\bm R_2$, allowing its macroscopic form to be learned independently.

For the training, we include a loss term with anchoring that fixes the entropy at equilibrium $s^\NN_0(c_1=1,c_2=1)=0$:
\begin{align}
  \mathcal{L}_{\sigma}
  &= \mathbb{E}\Big[
      \sum^d_{\alpha=1}w_\alpha \,\big(\,2\hat c_\alpha\,\sigma_\alpha^{\NN}+\hat\tau_\alpha\big)^2
    \Big]
    + w_0\big(s^{\NN}_0\big)^2,
  \label{eq:stage1_loss}
\end{align}
where $\hat c_\alpha$ are the computed eigenvalues of the data \(\hat {\bm c}\),
and the terms in parentheses compare the measured stress eigenvalues
\(\hat\tau_\alpha\) with the ones reconstructed from the GENERIC relation
using \(\sigma_\alpha^{\NN}(\hat c)\) and the eigenvector decomposition
shared with \(\hat {\bm c}\). In the numerical examples, the arbitrary choice of loss-function weights is eliminated by setting 
$w_1=w_2=1$. Consequently, the relative contribution of the different configurations is determined solely by the point sampling described above. The anchor term fixes the entropy gauge at \(s^{\NN}({\bm 1})=0\), with a weight $w_0$ equal to $10^{-4}$.

In an initial step, the eigenvalues $\hat c_\alpha$ and $\hat \tau_\alpha$ with $\alpha=1,2$ {(in 2D)} are computed from the conformation tensor $\hat {\bm c}$ and the stress $\hat {\bm \tau}$ data. Then, a multilayer perceptron (MLP) with 64 neurons
and three hidden layers and \(\tanh\)
activation functions is trained. Once the network is trained, automatic differentiation can be used to compute the eigenvalues $\sigma_\alpha^\NN$ from 
the neural network entropy potential $s_c^{\NN}$ 
%using its definition in Eq. (\ref{eq:sigma_OB}). 
The tensor system is reconstructed
on a physical basis using the $\hat {\bm c}$ eigenvectors and Eq. (\ref{eq:diag}) (see Appendix~\ref{sec:GENERIC_formulation}).
Finally, the neural-network training is performed with an Adam solver using a learning rate of
\(10^{-2}\) with reduction on plateau and a weight decay of
\(10^{-5}\). Training convergence required around \(10^5\) epochs.

\subsubsection*{Giesekus training example for Stage 1}
\label{sec:stage1_results}

We first run a {standard finite-volume Rheotool} simulation of the Giesekus flow around a cylinder at {\(\Wi=5.0\) (Figure \ref{fig:geometries} - left) with \(\alpha=0.3\), $\rho=1$ and constant $\eta_s=0.59,\, \eta_p=0.41$}. The simulation provides \(\hat {\bm c}\), \(\hat {\bm \tau}\), $\hat{\mathbf{v}}$, and \(\widehat{\nabla \bm v}\)
at each FV mesh point. Then \(\hat {\bm c}\) and \(\hat {\bm \tau}\) datasets are used to train the first neural network ${\bm \sigma}^\NN$ following the procedure described in Section~\ref{subsec:stage1}. 

\begin{figure}[hbt!]
\begin{center}
    \includegraphics[width=1.0\linewidth]{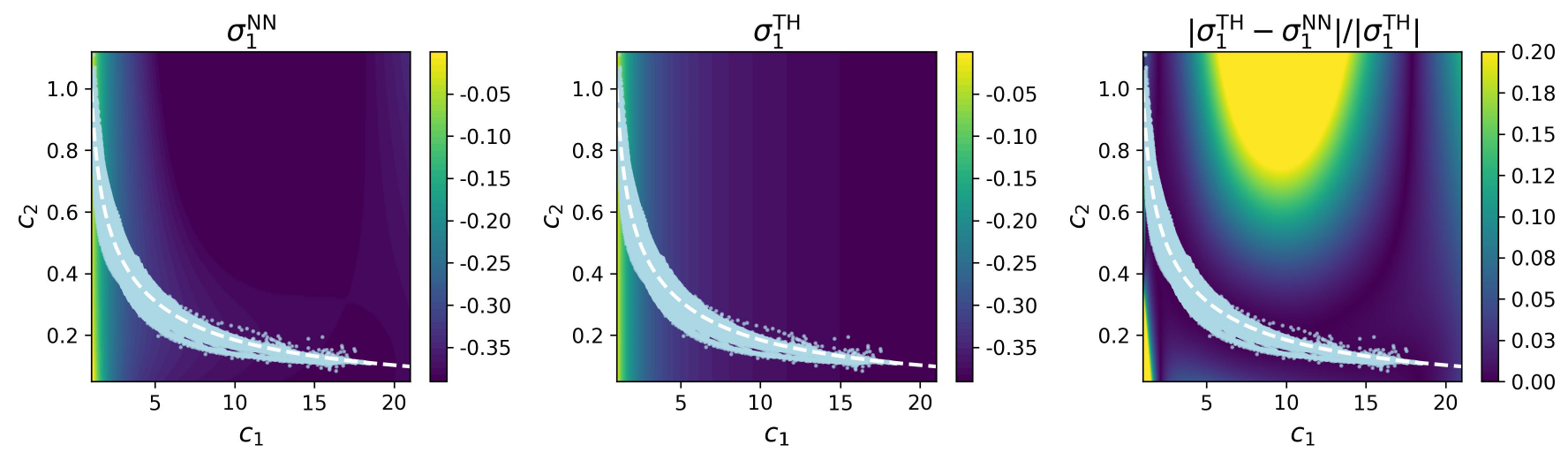}
    \includegraphics[width=1.0\linewidth]{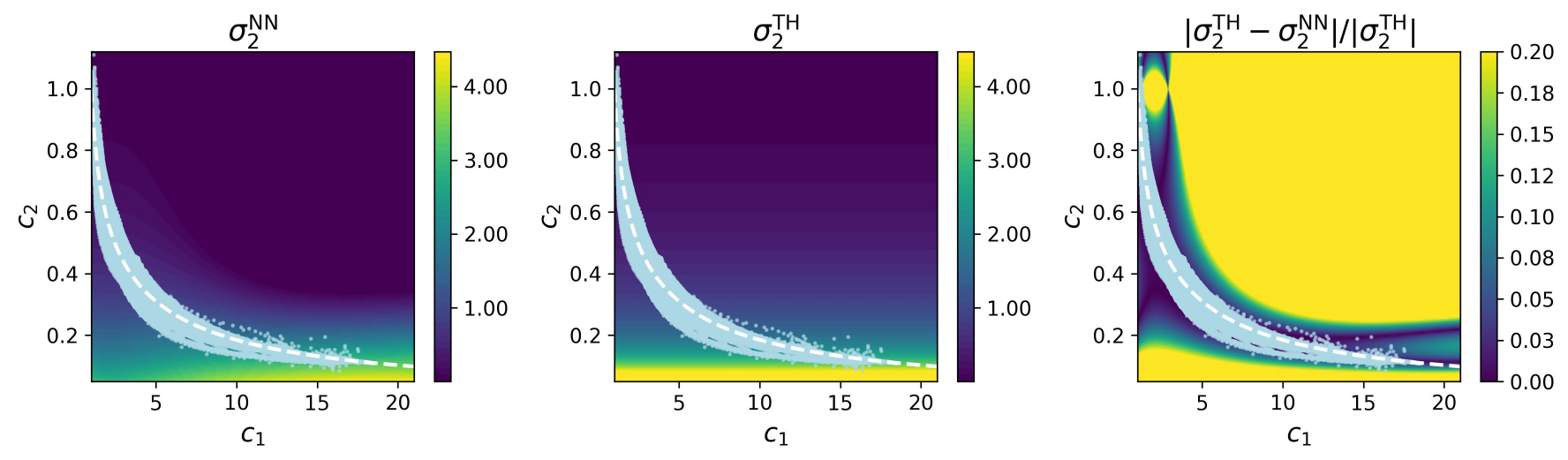}
\end{center}
\caption{Stage~1 training results for flow around a cylinder with $\Wi=5.0$. The white dashed line represents the steady-state rheometric flow. The light blue dots represent the training points for the eigenvalues of $\bm c$.}
\label{fig:training_results_stage1_wi5p0}
\end{figure}

Stage 1 training results for $\Wi=5.0$ are shown in
Figure~\ref{fig:training_results_stage1_wi5p0}: a comparison of the eigenvalues of entropy conjugate variable $\bm \sigma$ {in the eigenspace} for the NN approximator $\sigma^\NN_\alpha$ (first column) and the Giesekus model $\sigma^\TH_\alpha$ (second column). Additionally, in the third column the relative error of the approximator predictions is presented. It is known that such viscoelastic steady-state planar extensional flow and simple shear follow an identical trajectory in the $c_1$–$c_2$ space \cite{Nieto2023,Nieto2025}. The points of that trajectory, the steady-state rheometric response, are represented for the Giesekus case by the white dashed line of the figures. The light blue dots represent the training points for the eigenvalues of $\bm c$ for flow around a cylinder at Wi=5. We observe the relative error to be excellent ($<2.5$ \%) in the region covered by the training points. The error in $\sigma_1$ can become larger in the region with large $c_1$ values outside the training range, but in any case the method is generally good at predicting values that are far from the training region. For $\sigma_2$, the relative error can become quite significant in regions outside but close to the training region.  

\subsection{Stage 2: learning the friction matrix}
\label{sec:stage2}

Stage~2 learns ${\bm R}_2^\NN$ by minimizing the conformation-tensor evolution
residual with measured kinematics $\hat {\bm u}$ and $\widehat{ \nabla {\bm u}}$ and the conformation tensor data $\hat {\bm c}$. 

\begin{equation}
  {\bm e}_c
  := \ucd{\hat {\bm c}}
  - \bm R_2^{\NN}(\hat {\bm c})\,{\bm \sigma}^{\NN}(\hat {\bm c}),
  \qquad
  \mathcal{L}_{R_2}
  = \mathbb{E}\big[\|{\bm e}_c\|^2\big].
  \label{eq:stage2_residual}
\end{equation}
where $\|\cdot\|^2$ denotes the squared Frobenius norm. Equivalently, working with the eigenvalues according to Eq. (\ref{eq:eigenvalue_G}), we consider the decoupled
residual components
\begin{equation}
  e_{c,\alpha} = \ucd{\hat c_\alpha} - R_{2,\alpha}^{\NN}(\hat c)\,\sigma^{\NN}_\alpha(\hat c),
  \qquad
   \mathcal{L}_{R_2}
  = \mathbb{E}\big[ \sum_{\alpha=1}^d e_{c,\alpha}^2\big],
  \label{eq:stage2_eigen_residual}
\end{equation}
where \(R_{2,\alpha}^{\NN}\) and \(\sigma_\alpha^{\NN}\) are the friction matrix and
entropy-conjugate eigenvalues evaluated at \(\hat c = \{\hat c_1, \hat c_2\}\). Note that $\alpha$ runs from 1 to $d$, the dimensionality of the problem. In our approach, for simplicity we have considered 2D problems ($d=2$) but the extension of the methodology to 3D is straightforward. The residual makes use of the Stage~1 approximator for the predicted values of \(\sigma^\NN(\hat c)\). Note that the loss function is
purely based on the residual of the evolution equation for $\bm c$ (i.e., there is no supervised target for \( R_{2,\alpha}^{\NN}\) once $\sigma_\alpha^\NN$ has been correctly learnt through stage 1). The training of the friction matrix eigenvalues \( R_{2,\alpha}^{\NN}\) uses the same architecture as Stage~1 with Adam solver (learning rate \(10^{-2}\), a weight decay \(10^{-5}\)) and a reduction on plateau scheduler over multiple rounds. 

As in Stage~1, ${\bm R}_2^\NN$ shares eigenvectors with the
conformation tensor allowing us to train a neural network for each of the eigenvalues. Accordingly, during
training and simulations, $R_{2,1}^\NN$ and $R_{2,2}^\NN$ are treated as independent neural networks. However, the 
eigenvalues ($R_{2,1}^\NN$ and $R_{2,2}^\NN$) are not restricted to depend exclusively on $c_1$
and $c_2$ separately. Our approach considers non-trivial complex functions of both eigenvalues of $\bm c$, that is $R_{2,1}(c_1,c_2)$ and $R_{2,2}(c_1,c_2)$. This complexity is to be expected since  traditional constitutive modeling has often relied on the use of invariants of the conformation tensor such as ${\rm tr}(\bm c)$ or ${\rm det}(\bm c)$.

\subsubsection*{Giesekus training example for Stage 2}
\label{sec:stage2_results}

\begin{figure}[hbt!]
\begin{center}
    \includegraphics[width=1.0\linewidth]{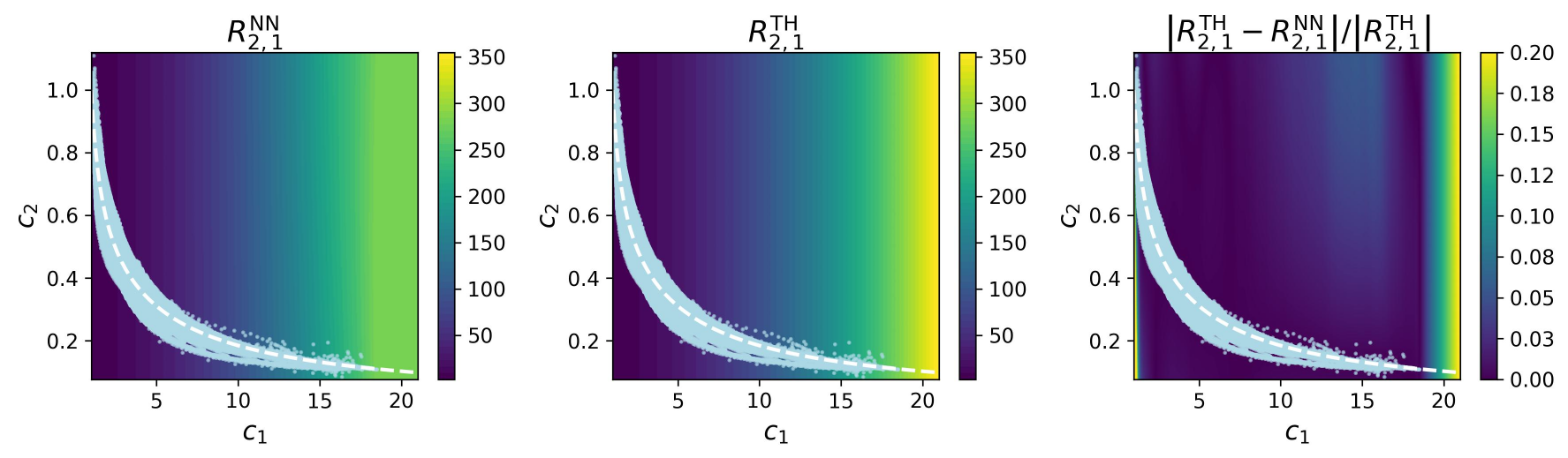}
    \includegraphics[width=1.0\linewidth]{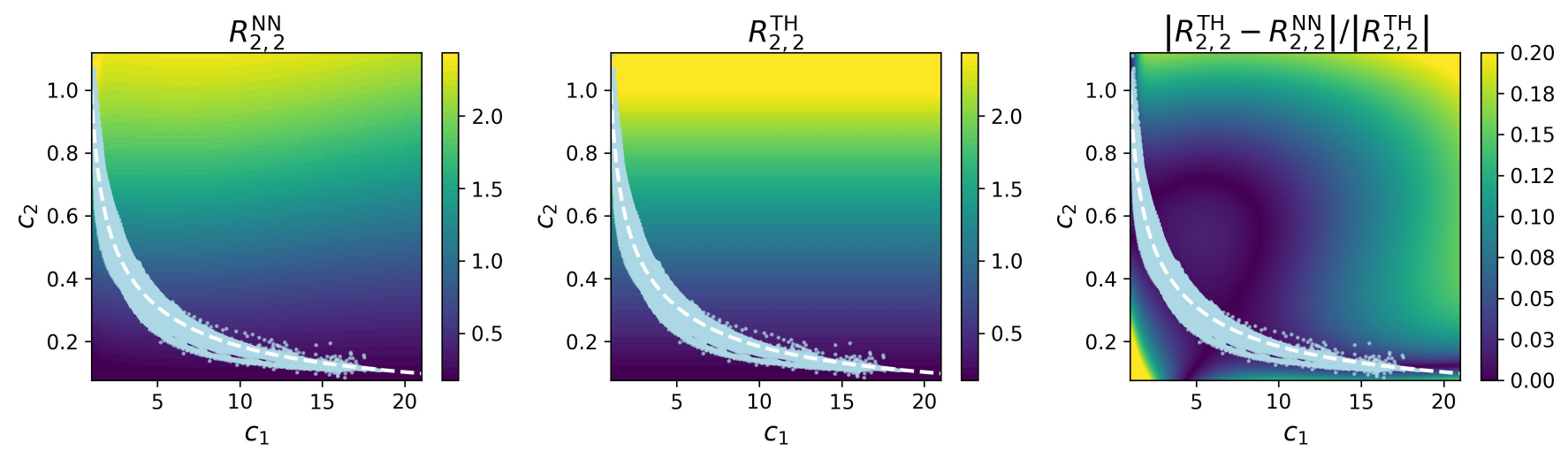}
\end{center}
\caption{Stage~2 training results for flow around a cylinder with $\Wi=5.0$. The white dashed line represents the steady-state rheometric flow. The light blue dots represent the training points for the eigenvalues of $\bm c$.}
\label{fig:training_results_stage2_wi5p0}
\end{figure}

Following the procedure in the preceding section,%Section \ref{subsec:stage2}
 the neural network for the friction matrix ${\bm R_2}^\NN$ is trained taking
{\(\hat {\bm c}\), $\hat{\mathbf{v}}$, and \(\widehat{\nabla \mathbf v}\)}
data from the same finite-volume simulation of the Giesekus flow around a cylinder at {\(\Wi=5.0\)} with \(\alpha=0.3\) also used in Stage 1 training. 

Figure~\ref{fig:training_results_stage2_wi5p0} presents a comparison between the {ground-truth} Giesekus model (TH) and the neural network approximator (NN) for the eigenvalues of the friction matrix ${\bm R}_2^\NN$. The white dashed line represents the steady-state rheometric flow for the Giesekus model (i.e., steady-state extensional or simple shear). The light blue dots represent the training points for the eigenvalues of $\bm c$ for flow around a cylinder at Wi=5. As in the case of Figure~\ref{fig:training_results_stage1_wi5p0}, predictions are excellent in the training region. However, in the case of ${\bm R}_2^\NN$ the relative error stays bounded in a broad region outside the training space. The most significant deviations from the Giesekus model are observed once the maximum $c_1$ in training is exceeded. This result is significant since the large errors in the predictions for $\sigma_2$ and $R_{2,1}$ outside the training domain are the key limitation for the extrapolation ability of the GINNGER approach.

\section{Numerical results}
\label{sec:results}

\begin{figure}[t]
    \centering
    \begin{center}
        \includegraphics[width=0.55\linewidth]{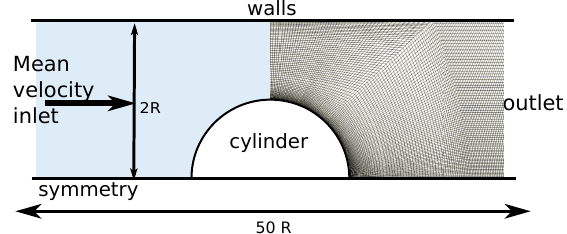}
        \includegraphics[width=0.4\linewidth]{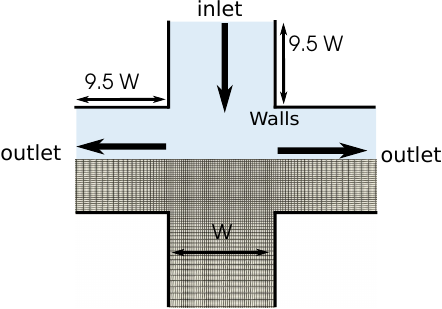}
    \end{center}
    \caption{Sketches of the geometries/meshes used in the numerical Rheotool examples: flow around a cylinder (left) and cross-slot (right).}
    \label{fig:geometries}
\end{figure}

In the remainder of the manuscript, we will continue to use the Giesekus model as an example constitutive model with non-trivial $\bm \sigma$ and $\bm R_2$. Three numerical tests are used to evaluate the performance of the GINNGER procedure, each with a different illustrative purpose. As test cases we use a classical flow around a cylinder and a cross-slot flow. The geometries are shown in Figure \ref{fig:geometries} (left and right panels respectively). The Weissenberg number is defined as $\Wi = \lambda V/L$, where $V$ is the mean velocity in the constricted gap between the cylinder and the channel wall and $L$ is a geometry-specific reference length---the cylinder radius $R$ for the flow-around-a-cylinder benchmark (blockage ratio $0.5$, following Alves et al.~\cite{Alves2001}) and the channel width $W$ for the cross-slot geometry.

First, we validate our training by comparing the ground-truth Giesekus constitutive equation and the two-stage GINNGER closure  in the standard finite volume solution to flow-around-a-cylinder benchmark at $\Wi=5.0$ (i.e. we compare the GINNGER model with the training data in the same flow geometry and the same Wi number that was used for the training). 

Second, we test the ability of the learned model to be used at flow regimes other than the training in flow around a cylinder at Wi=5.0. For that, we run the numerical benchmark -- flow around a cylinder -- at $\Wi=1.0,3.0$ and $5.1$. This covers a range of Wi from well below the training regime up to slightly above (i.e. flow geometry remains the same but flow rates are modified). 

Finally, we check model transferability, i.e. whether the GINNGER learned model performs accurately out-of-distribution, i.e. across geometries different from those used in training. To that end, we transfer the GINNGER model trained in flow around a cylinder at $\Wi=5.0$ to a cross-slot flow {\bf without any re-training}, thereby demonstrating the geometry-agnostic character of the method.

\subsection{GINNGER validation: flow around a cylinder}
\label{sec:results_cylinder}

In this section, the benchmark simulation for flow around a cylinder at Wi=5.0 is performed with the constitutive model provided by the neural-network constitutive {approximation} (${\bm \sigma}^\NN$ and ${\bm R}_2^\NN$). We validate the procedure by
comparing results from the traditional finite-volume solver using the Giesekus model and the hybrid two-stage neural-network/finite-volume approach.

\begin{figure}[hbt!]
  \centering
  \begin{center}
    \includegraphics[width=0.4\linewidth]{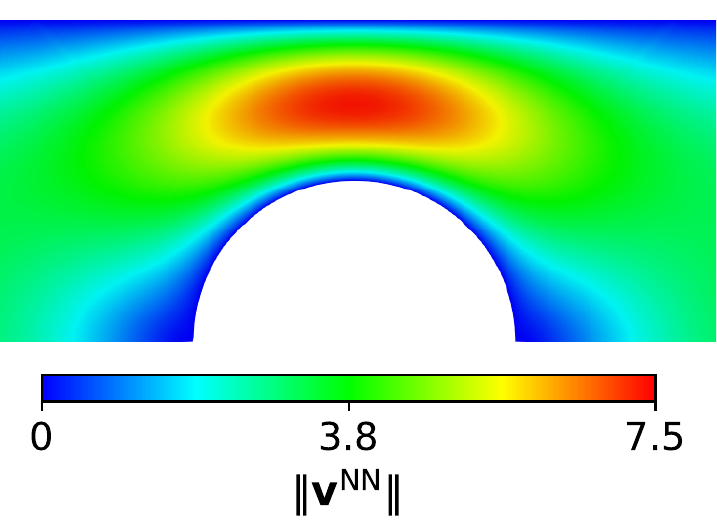}
      \includegraphics[width=0.4\linewidth]{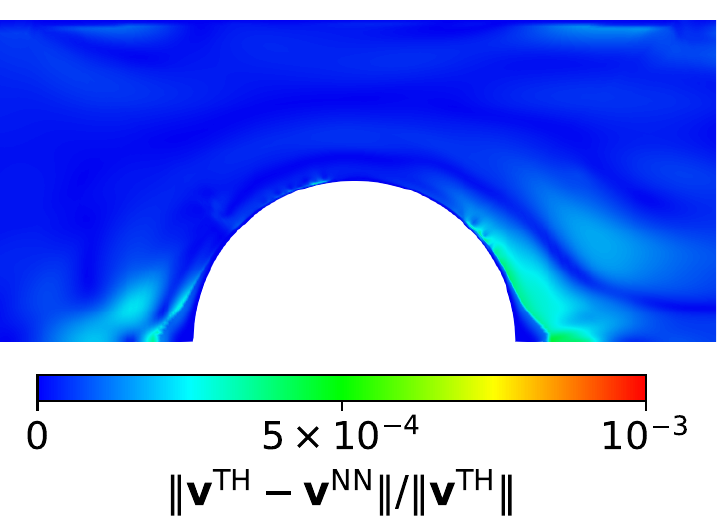}
      \includegraphics[width=0.4\linewidth]{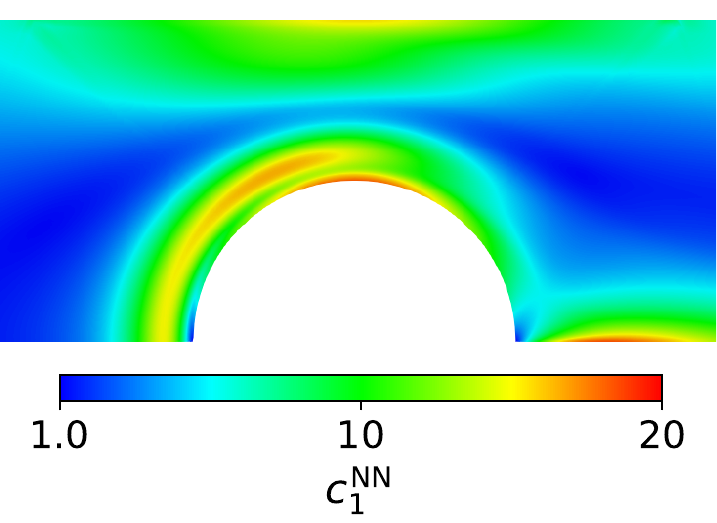}
      \includegraphics[width=0.4\linewidth]{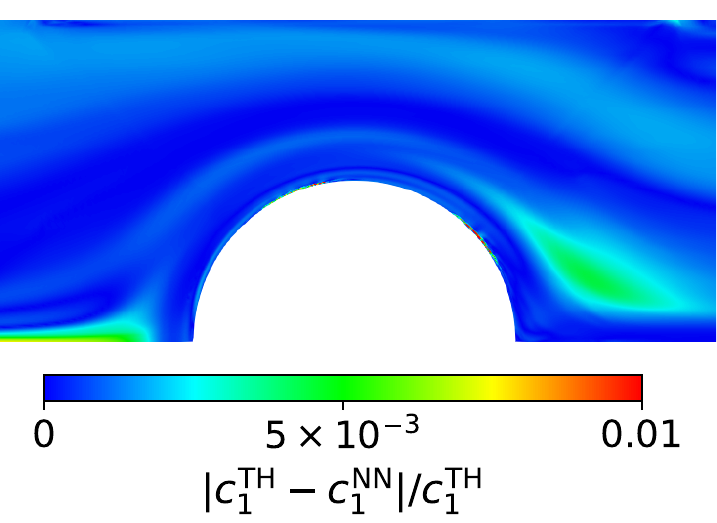}
      \includegraphics[width=0.4\linewidth]{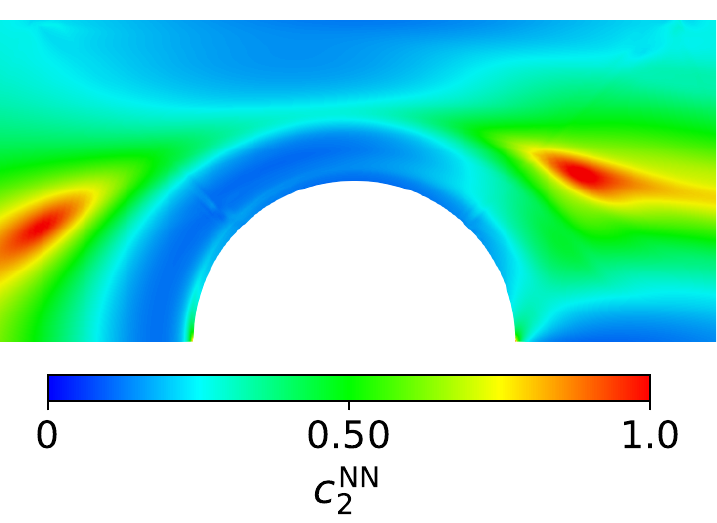}
      \includegraphics[width=0.4\linewidth]{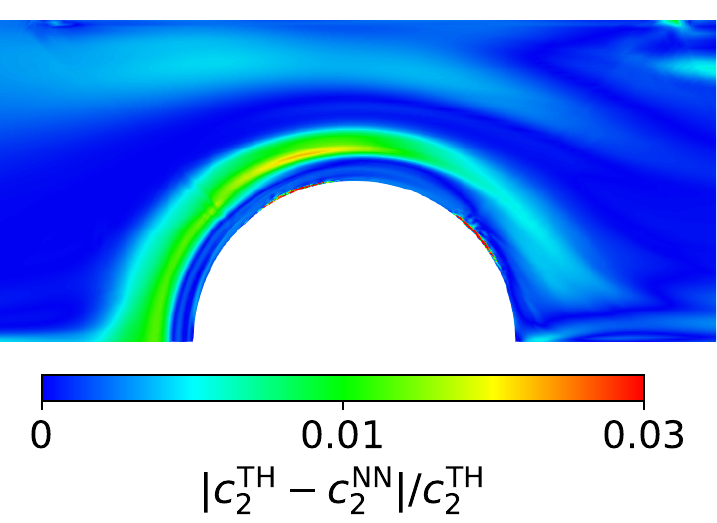}
      \includegraphics[width=0.4\linewidth]{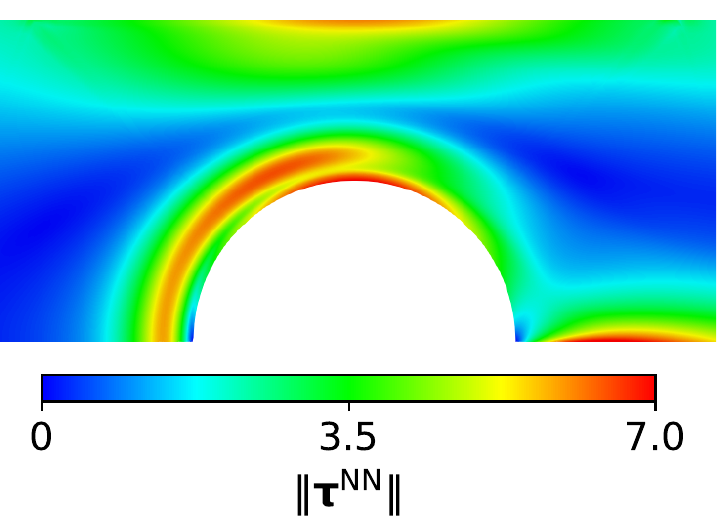}
      \includegraphics[width=0.4\linewidth]{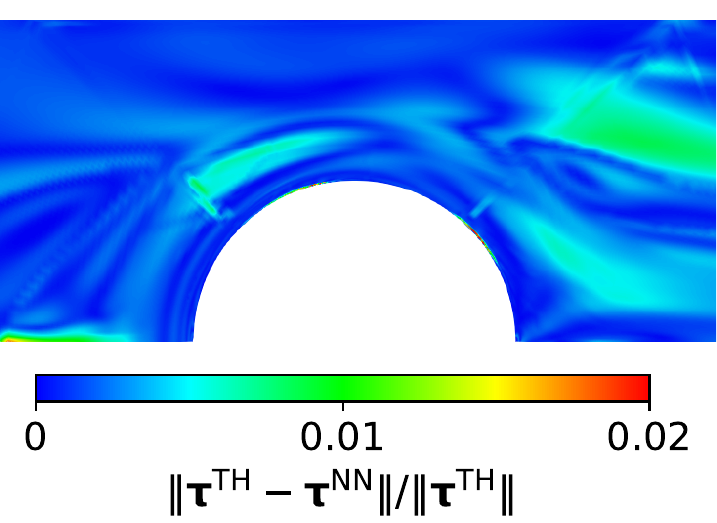}
  \end{center}
\caption{GINNGER-FV simulation results for $||\mathbf{v}||$, $c_1$, $c_2$, and $||\bm \tau||$ fields (left) and their relative-error fields (right) in flow around a cylinder at $\Wi=5$ {(training regime)}.}
  \label{fig:wi5p0_fields}
\end{figure}

In Figure \ref{fig:wi5p0_fields} we
show the most relevant fields in the simulations ($||\mathbf{v}||$, $c_1$, $c_2$, and $||\bm{\tau}||$) together with their relative-error fields in physical space. 
Throughout this work, the 
magnitudes of the velocity and stress fields (i.e. $v$ and $\tau$ respectively) are evaluated using the Frobenius norm. The relative error at each point is computed by dividing the error tensor Frobenius magnitude by the reference solution magnitude. To prevent the relative error from diverging when the reference field ($|{\rm ref}|$) approaches zero (e.g.\ near equilibrium or at stagnation points), an exponential damping factor of $\exp(0.05 \cdot G_{\rm max}/|{\rm ref}|$) is applied when the reference magnitude falls below 5\% of its global maximum ($G_{\rm max}$), driving the reported error toward zero in near-equilibrium regions.

\FloatBarrier

\begin{figure}[t]
    \begin{center}
    \includegraphics[width=0.8\linewidth]{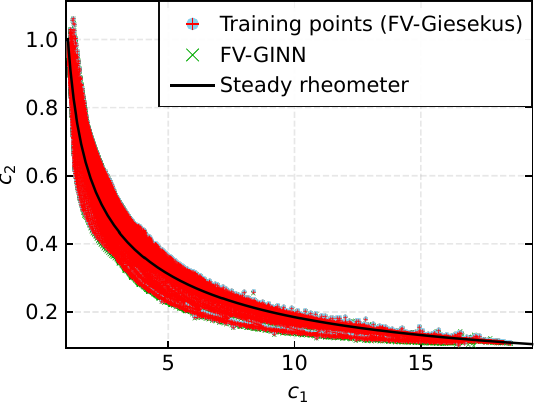}
    \end{center}
  \caption{{Steady-state} $c_1$-$c_2$ maps for the flow-around-cylinder validation at $\Wi=5.0$. %Simulation results are given by: FV-Giesekus (\textcolor{red}{$+$}) and FV-GINNGER simulation (\textcolor{green}{$\times$}). The solid black line represent steady-state rheometric flow.
  }
  \label{fig:c1_c2_map_cylinder_Wi5.0}
\end{figure}

This way, the regions where these variables have very small magnitude do not dominate the colormap while finite-stress regions remain clearly visible. The damped relative-error field plots (e.g. Figure  \ref{fig:wi5p0_fields}) show
the local error distribution once those near-zero regions are de-emphasized; the errors remain small throughout the domain, including in the highly stretched wake where the effects of the learned closure are most relevant. {Some higher relative errors are observed close to the boundary of the cylinder. These originate from a reduction in the GINNGER accuracy as we approach the maximum value of $c_1$ present in the training data (see Figure \ref{fig:c1_c2_map_cylinder_Wi5.0}).

Figure \ref{fig:c1_c2_map_cylinder_Wi5.0} presents the $c_1$-$c_2$ map for the FV simulations using GINNGER}, the FV simulations using Giesekus (FV) together with the steady-state rheometric line (i.e.  the analytic solution for steady-state extensional or shear flow). Note that in this case the points for the FV-Giesekus simulation $(c_1,c_2)$ correspond with the training data points. 
The agreement between both closure forms -- ground-truth Giesekus and GINNGER -- is excellent, a first result that validates our methodology.

Finally, following Sunol et al.~\cite{Sunol2025}, we
report a global Relative Root Mean Square Error (RRMSE) for each variable in Table \ref{tab:rrmse_all}. This metric remains
finite near equilibrium, where point-wise relative errors would diverge
as the stresses become very small, while still condensing each field to
a single scalar. Table \ref{tab:rrmse_all} shows the RRMSE values for all the simulations discussed in the manuscript (i.e. flow around a cylinder for a range of Wi and Cross-slot at Wi=0.5 { in the next section}). The RRMSE values for flow around a cylinder at the same Wi as in the training (Wi=5) stay well below 1 \%.  

\begin{table}[hbt!]
\centering
\caption{{Global RRMSE values for all test cases.}}
\label{tab:rrmse_all}
\footnotesize
\setlength{\tabcolsep}{2pt}
\begin{ruledtabular}
\begin{tabular}{@{}lccccc@{}}
\colrule
\shortstack{Variable\\$\times 10^{-5}$} & Cylinder & Cylinder & Cylinder & Cylinder & \shortstack{Cross-\\slot} \\
 & $\Wi=1.0$ & $\Wi=3.0$ & $\Wi=5.0$ & $\Wi=5.1$ & $\Wi=0.5$ \\
\colrule
$\|\mathbf{v}\|$ & 27.9 & 11.8 & 8.07 & 13.2 & 5.98 \\
$c_1$ & 351 & 113 & 253 & 380 & 105 \\
$c_2$ & 348 & 189 & 264 & 325 & 345 \\
%$\tau_1$ & $3.14\times 10^{-3}$ & $1.32\times 10^{-3}$ & $2.82\times 10^{-3}$ & $4.22\times 10^{-3}$ & $1.65\times 10^{-3}$ \\
%$\tau_2$ & $8.37\times 10^{-3}$ & $4.94\times 10^{-3}$ & $1.00\times 10^{-2}$ & $6.02\times 10^{-3}$ & $6.55\times 10^{-3}$ \\
$\|\bm{\tau}\|$ & 217 & 109 & 277 & 416 & 146 \\
\colrule
\end{tabular}
\end{ruledtabular}
\end{table}

\subsection{GINNGER extension: flow around a cylinder at different Wi numbers}
\label{sec:results_wi_extension}

In this section, we demonstrate the ability of the method to adapt to flow regimes for which it was not explicitly trained. We consider the flow
around a cylinder with $\Wi=1.0, 3.0$, and $5.1$ while using the neural-network model
trained only at $\Wi=5$. 

\begin{figure}[hbt!]
  \centering
  \begin{center}
    \includegraphics[width=0.4\linewidth]{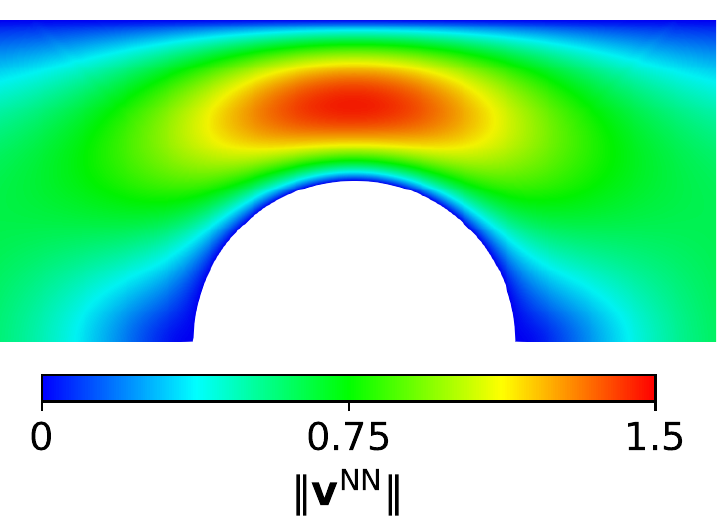}
      \includegraphics[width=0.4\linewidth]{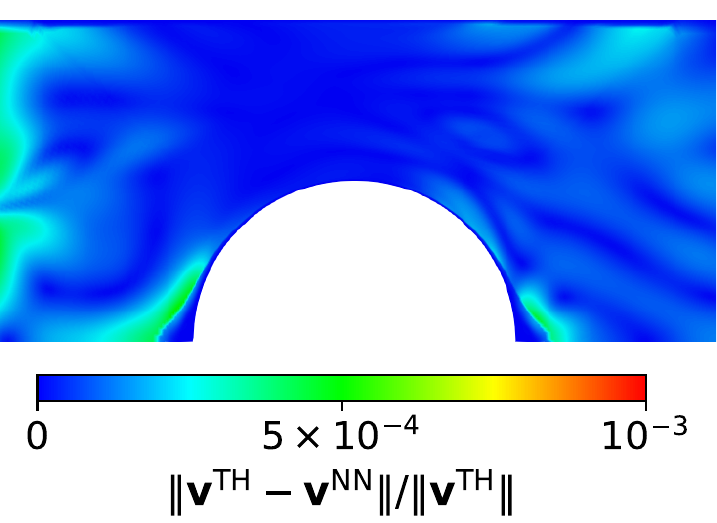}
      \includegraphics[width=0.4\linewidth]{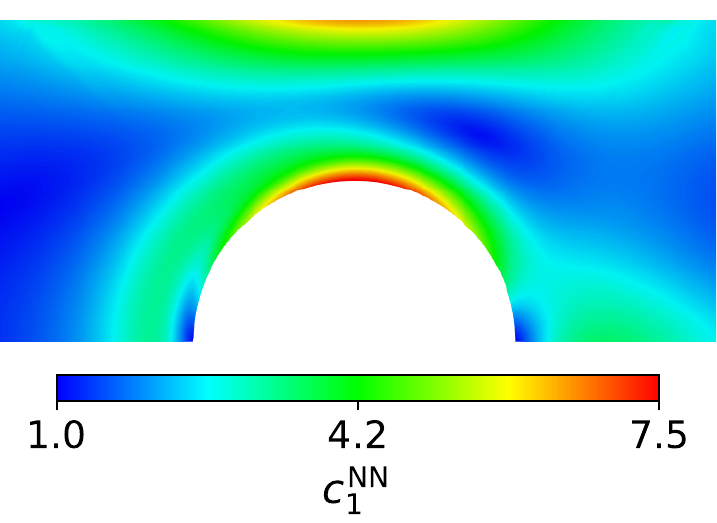}
      \includegraphics[width=0.4\linewidth]{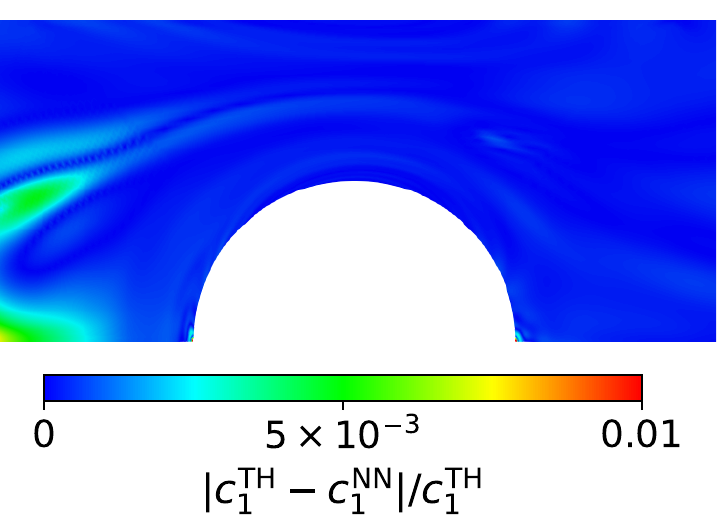}
      \includegraphics[width=0.4\linewidth]{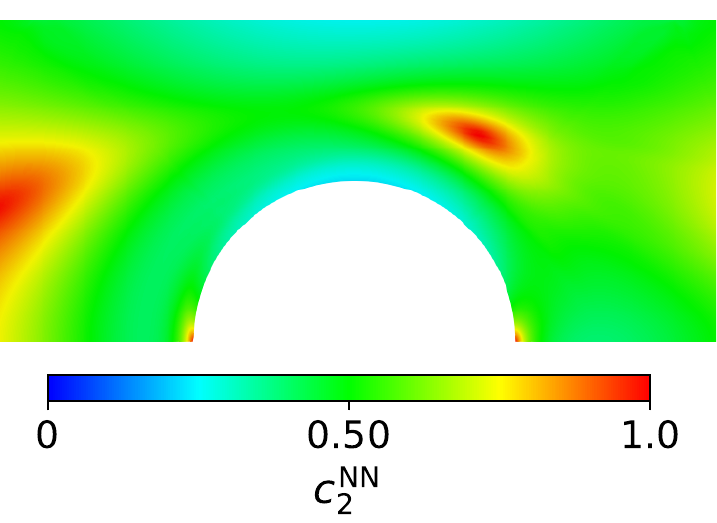}
      \includegraphics[width=0.4\linewidth]{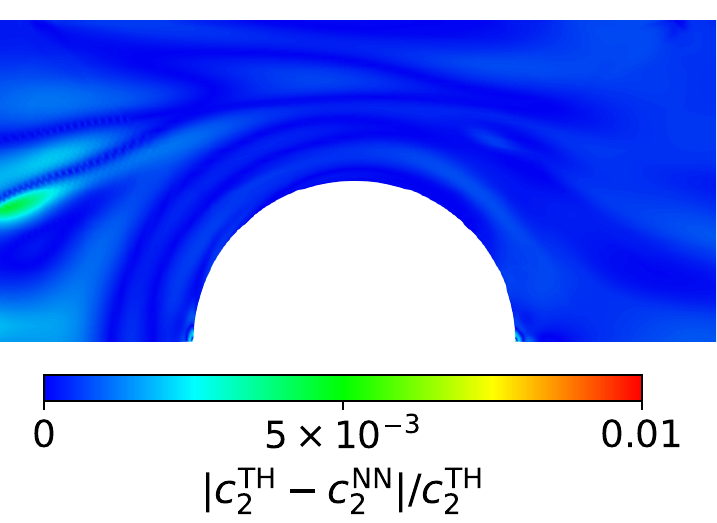}
      \includegraphics[width=0.4\linewidth]{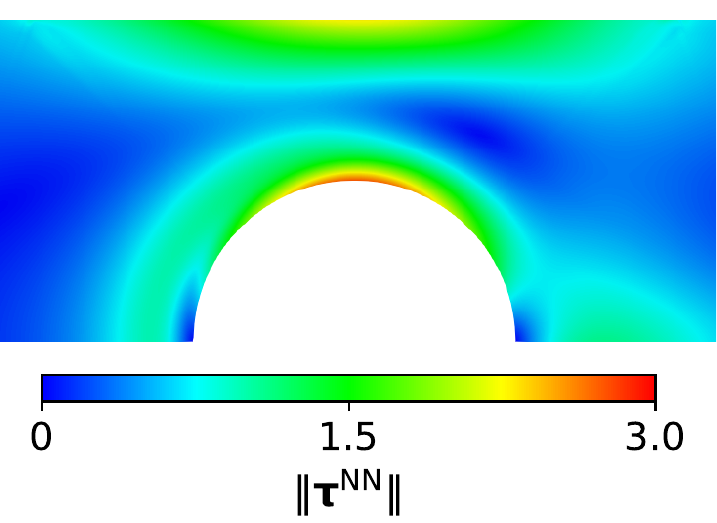}
      \includegraphics[width=0.4\linewidth]{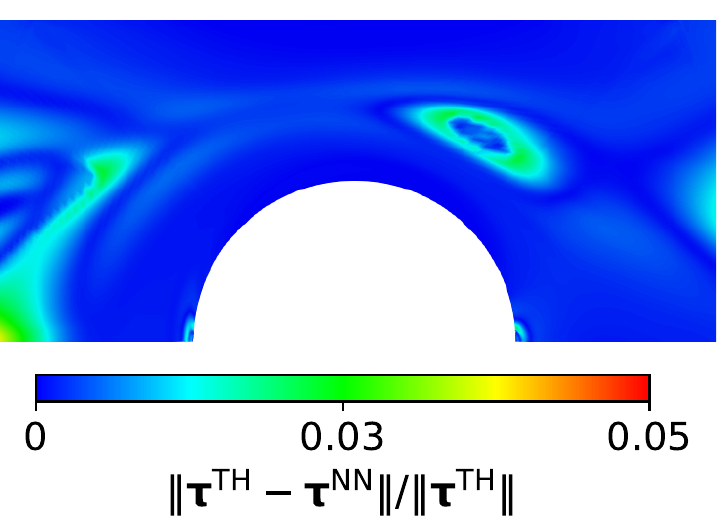}
  \end{center}
\caption{GINNGER-FV simulation results for $||\mathbf{v}||$, $c_1$, $c_2$, and $|| \bm \tau||$ fields (left) and their relative-error fields (right) in flow around a cylinder at $\Wi=1$.}
  \label{fig:wi1p0_fields}
\end{figure}

\begin{figure}[t]
  \centering
  \includegraphics[width=0.8\linewidth]{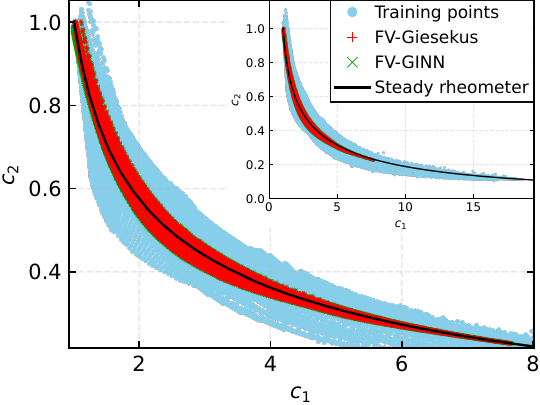}
  \caption{$c_1$-$c_2$ maps for the flow-around-cylinder validation at $\Wi=1.0$. Blue points correspond to the $c_1-c_2$ map trained at $\Wi=5$.
  %Simulation results are given by: FV-Giesekus (\textcolor{red}{$+$}) and FV-GINNGER simulation (\textcolor{green}{$\times$}). The solid black line represent steady-state rheometric flow. The light blue points line correspond to the training data.
  }
  \label{fig:c1_c2_map_cylinder_Wi1.0}
\end{figure}

Figure \ref{fig:wi1p0_fields} shows the representative simulation results fields $||\mathbf{v}||$, $c_1$, $c_2$, and $|| \bm \tau||$ for a GINNGER-FV simulation of flow around a cylinder at Wi=1.0. Analogously to Figure \ref{fig:wi5p0_fields}, in the right-side panels, in Figure \ref{fig:wi1p0_fields} we present the relative errors resulting from the comparison with a {ground-truth} Giesekus-FV simulation at Wi=1.0. Results are excellent with the highest error being observed for the stress and reaching only 1\% in the regions where stress is expected to be near zero.  The explanation for the GINNGER model trained at Wi=5 being able to perform well at lower Wi for which it was not trained can be gathered from Figure \ref{fig:c1_c2_map_cylinder_Wi1.0}. The regions of the $c_1$-$c_2$ space explored at lower Wi are compatible with the training region ({light blue points}) where the accuracy of the PINNs approximation for $\sigma_\alpha^\NN$ and $R_{2,\alpha}^{\NN}$ has been shown to be excellent. 
Similar results can be observed for Wi=3.0 in Figures \ref{fig:wi3p0_fields} and \ref{fig:c1_c2_map_cylinder_Wi3.0} in Appendix \ref{app:app1}.

\begin{figure}[t]
  \centering
  \includegraphics[width=0.8\linewidth]{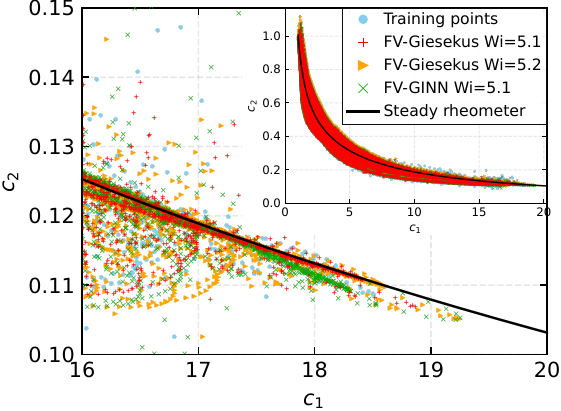}
  \caption{Zoom of the critical $c_1$-$c_2$ space for flow-around-cylinder validation at $\Wi > 5$. The inset shows the full $c_1$-$c_2$ map.}
  \label{fig:c1_c2_map_cylinder_Wi5.1}
\end{figure}

Next, we examine if the GINNGER model can perform accurate simulations at larger Wi than the one used in training. These conditions involve extrapolating outside the domain in the $c_1$-$c_2$ map used in training (see Figure \ref{fig:c1_c2_map_cylinder_Wi5.0}). In Figure \ref{fig:c1_c2_map_cylinder_Wi5.1} we present the $c_1$-$c_2$ space explored in: {(i)} simulations using FV-GINNGER and FV-Giesekus at Wi=5.1, {(ii)} FV-Giesekus at Wi=5.2 and {(iii)} the training data (i.e. FV-Giesekus at Wi=5.0). The FV-GINNGER simulations {numerically converge} up to Wi=5.1. The fields for $||\mathbf{v}||$, $c_1$, $c_2$, and $|| \bm \tau||$ are visually indistinguishable from those in Figure \ref{fig:wi5p0_fields} and therefore, are not presented. We ran simulations at Wi=5.2 that did not converge using FV-GINNGER, showing that the ``extrapolation'' ability of the GINNGER models is limited {outside the hull of training, an issue in PINNs already analyzed in \cite{Bonfanti2024a}}. For reference, the FV-Giesekus results for Wi=5.2 are presented in Figure \ref{fig:c1_c2_map_cylinder_Wi5.1}. Since $c_1$ is rapidly growing with Wi, a relatively small increase in Wi takes the explored region in the $c_1$-$c_2$ space to a region where the errors in $\sigma_\alpha$ and $R_{2,\alpha}$ are large. In particular, when the maximum values for $c_1$ grow very rapidly, the relative errors in $\sigma_2$ and $R_{2,1}$ grow rapidly {as well} (see Figures \ref{fig:training_results_stage1_wi5p0} and \ref{fig:training_results_stage2_wi5p0}), and {GINNGER predictions} are very sensitive to small fluctuations. 

In Figure \ref{fig:c1_c2_map_cylinder_Wi5.1} it can be observed that a number of points for FV-GINNGER at Wi=5.1 fall outside of the training hull (i.e. at $c_1\gtrsim 18.5$) with a degradation in results. This highlights one of the pillars on which this method relies: the natural state space of GINNGER is not geometry space but thermodynamic state given by the $c_1$-$c_2$ space, and it needs to interpolate within the spectral space spanned by the training data. It suffers from the extrapolation problems common to NN-based methods. From a practical perspective, this issue could be solved by training the GINNGER at increasingly larger Wi as needed but the limiting problem would still persist at higher Wi numbers.

Table \ref{tab:rrmse_all} {summarizes} the global Relative Root Mean Square Error (RRMSE) for $||\mathbf{v}||$, $c_1$, $c_2$, and $||\bm \tau||$ at Wi=1, Wi=3 and Wi=5.1. All errors stay below 1\%. 

\subsection{GINNGER transferability: planar cross-slot flow}
\label{sec:results_crossslot_tranfer}

Finally, we test the ability of the PINN trained in flow-around-cylinder at $\Wi=5.0$ to be transferred to cross-slot flow geometry. The cross-slot is an interesting test case since it presents both regions that are shear dominated and regions that are extension dominated. In particular, the center is often used as a test case for purely extensional flows \cite{Haward2016}. {As a test, Wi=0.5 is chosen in the cross slot flow.}
Note that the difference in Wi here is irrelevant since once the geometry is changed, {the flow character is also changed and} the Wi numbers cannot be compared {one-to-one}. As before, in this test we compare the FV-Giesekus model with the FV-GINNGER model.  

\begin{figure}[hbt!]
  \begin{center}
  \begin{tabular}{cc}
       \includegraphics[width=0.235\linewidth]{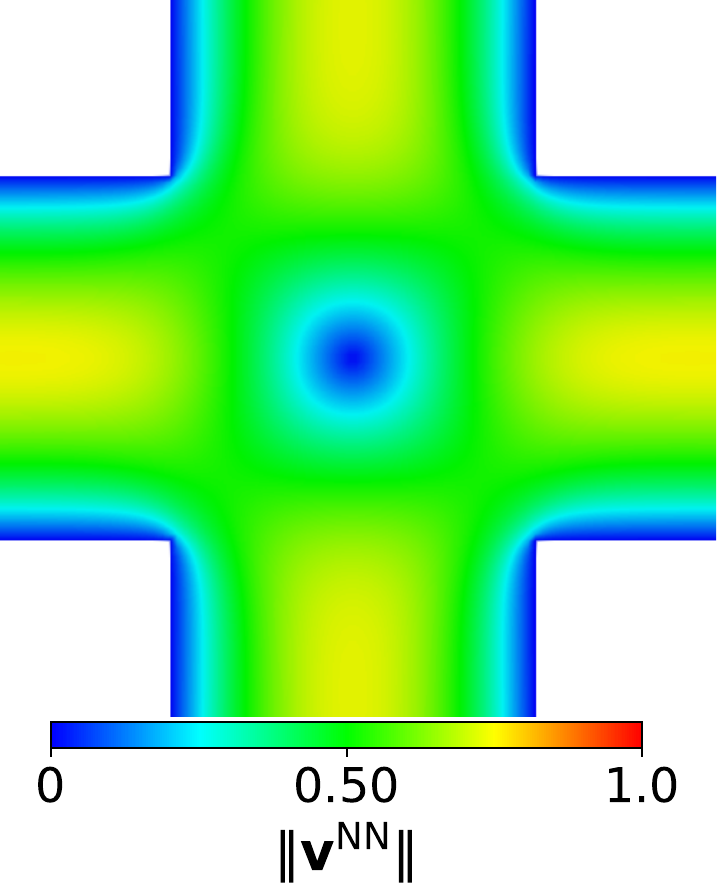} &  \includegraphics[width=0.235\linewidth]{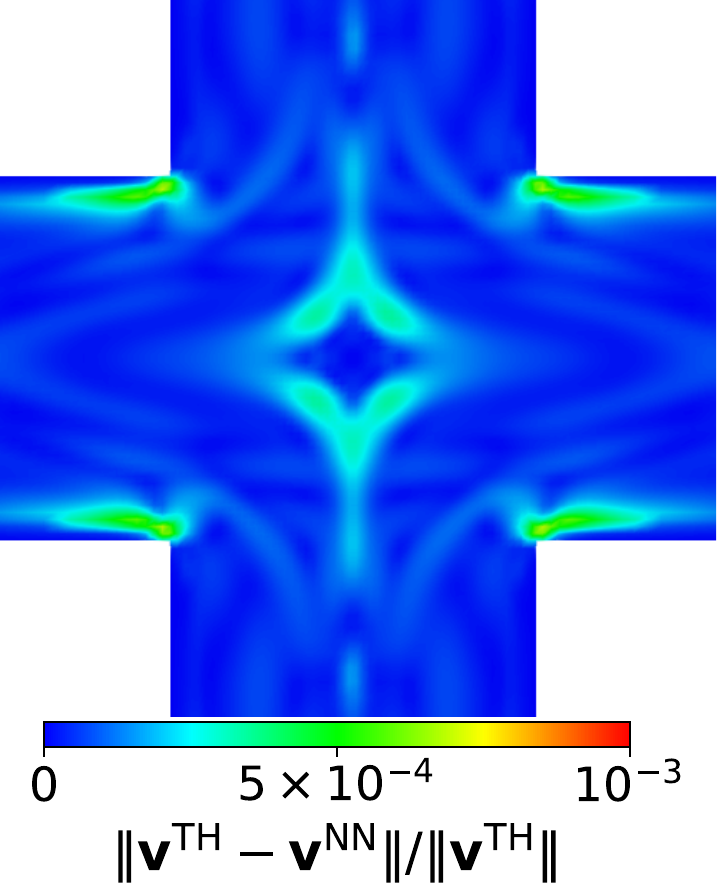}\\
       \includegraphics[width=0.235\linewidth]{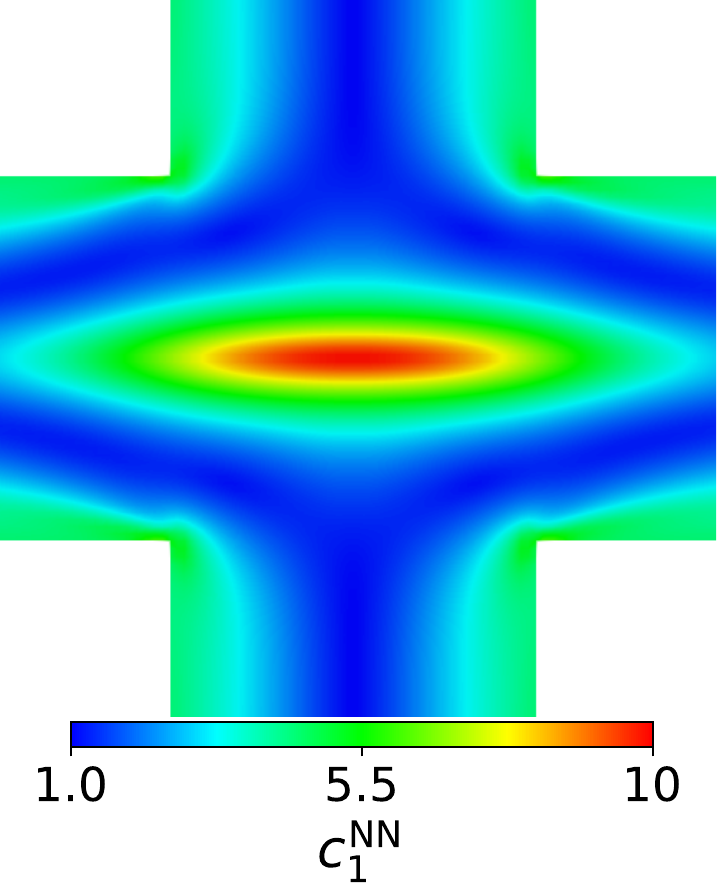} & \includegraphics[width=0.235\linewidth]{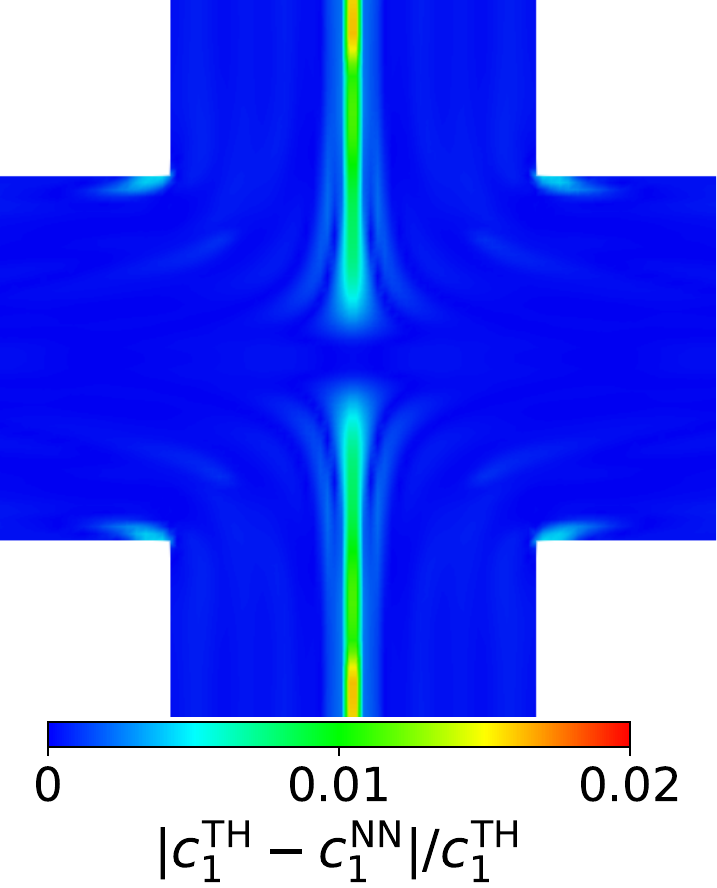}\\
       \includegraphics[width=0.235\linewidth]{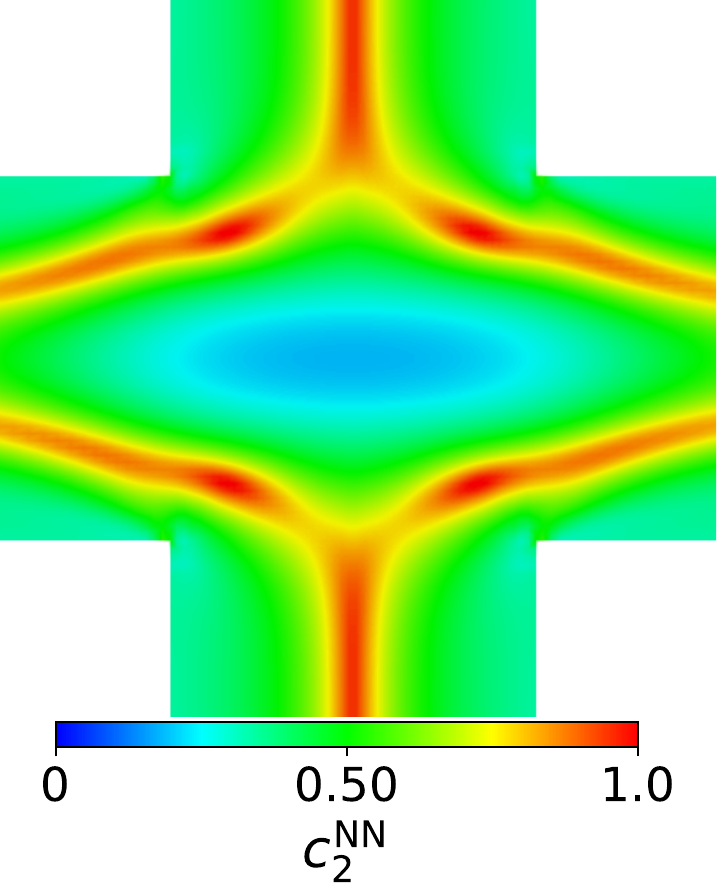} & \includegraphics[width=0.235\linewidth]{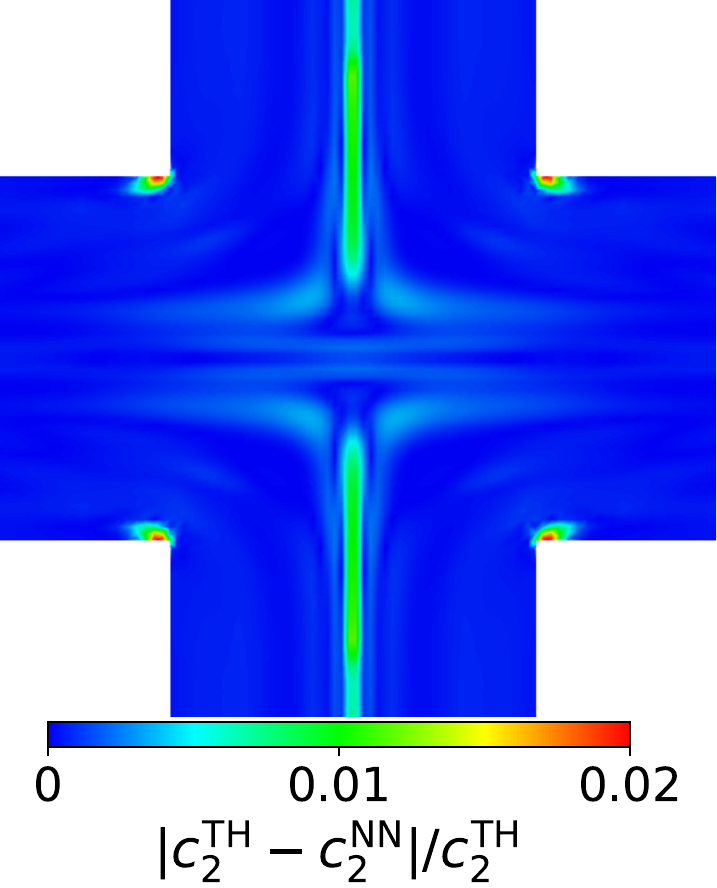}\\
       \includegraphics[width=0.235\linewidth]{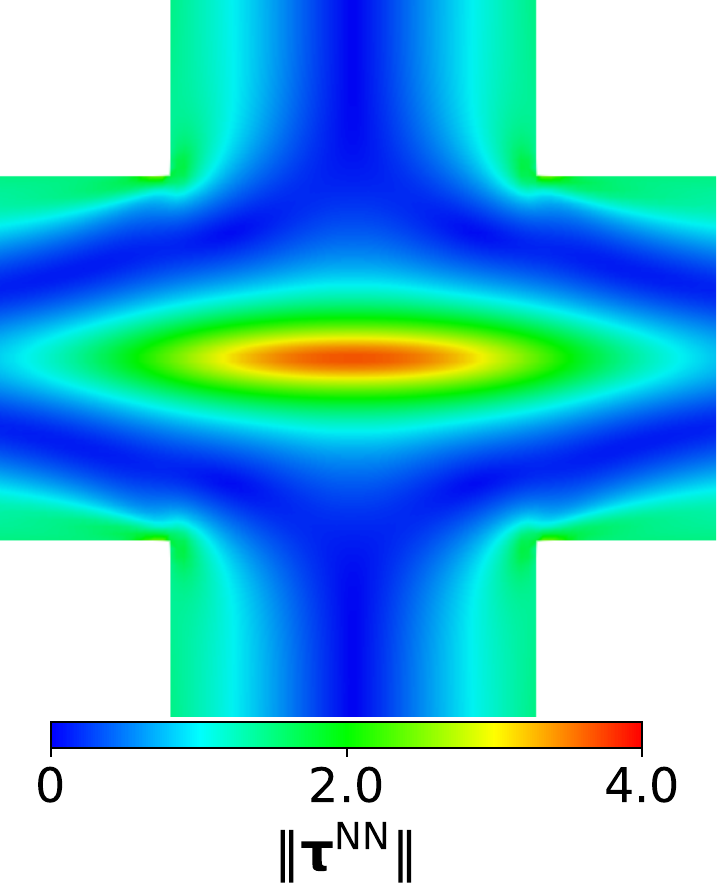} & 
        \includegraphics[width=0.235\linewidth]{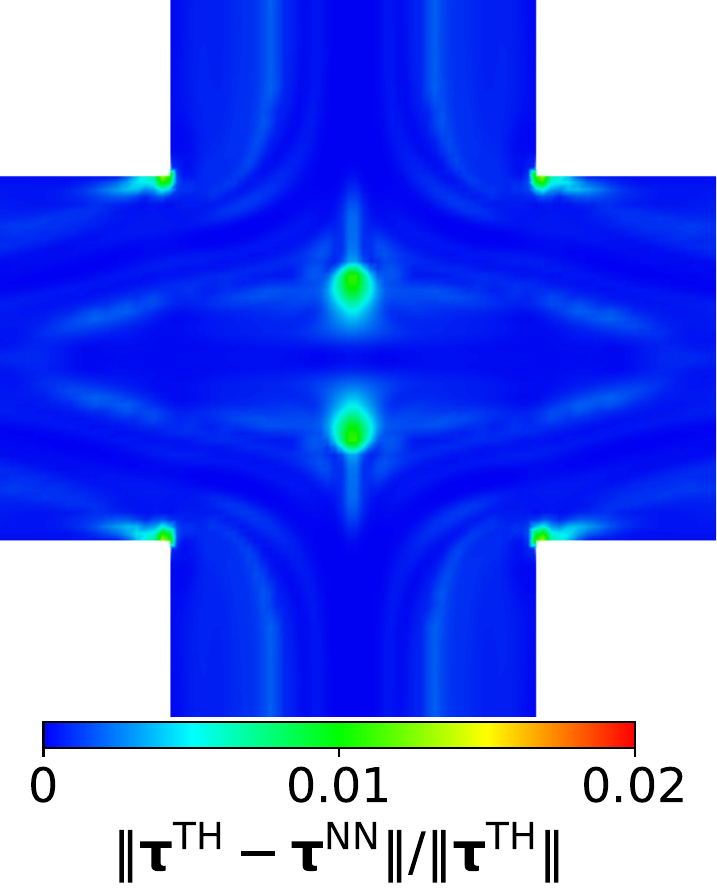} \end{tabular}
  \end{center}
\caption{GINNGER-FV simulation results for $\|\mathbf{v} \|$, $c_1$, $c_2$, and $\tau$ fields (left) and their relative-error fields (right) in cross-slot flow at $\Wi=0.5$.}
  \label{fig:wi0p5_crossSlot_fields}
\end{figure}

\begin{figure}[hbt!]
  \centering
    \includegraphics[width=0.8\textwidth]{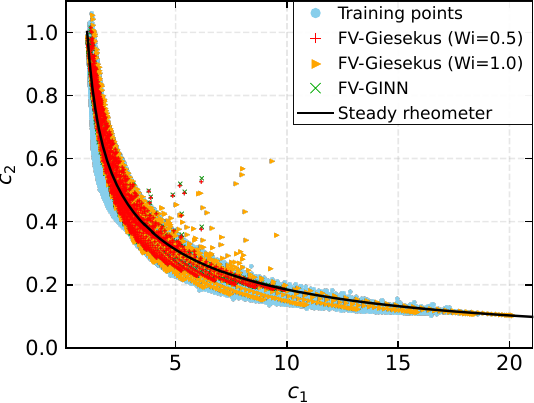} 
  \caption{$c_1$-$c_2$ map for cross-slot flow using FV-GINNGER model trained in flow around a cylinder at Wi=5.0.}
  \label{fig:c1_c2_map_crossslot}
\end{figure}

The representative fields ($||\mathbf{v}||$, $c_1$, $c_2$, and $||\bm \tau||$) resulting from the FV-GINNGER simulation and the damped relative error maps are
shown in Figure~\ref{fig:wi0p5_crossSlot_fields}. All relative errors are extremely low ($\lesssim 3$ \%) demonstrating that the GINNGER model performs well in a geometry very different from the one used for training. Furthermore, the highest error values are found in the sharp corners of the geometry. These can be attributed to {the singular stress values of the finite volume solution, which in the $c_1-c_2$ plane are represented by the isolated points around $(c_1,c_2) \approx (5,0.5)$ for this $\Wi=0.5$ case} (see Figure \ref{fig:c1_c2_map_crossslot}). These issues can be addressed with slightly modified ``smoother" geometries that are less prone to numerical errors \cite{Haward2016}. {It is also worth noticing that the highest error in $c_1$ and $c_2$ is found along the vertical symmetry line. In these regions the flow is shear dominated and because of said symmetry the deformation rate is close to zero. This results in $(c_1,c_2)\sim(1,1)$, where small variations are accentuated. Nevertheless, these regions of higher error do not strongly affect the overall simulation results since they are also characterized by vanishing $\bm \tau$, $\bm \sigma$ and $\bm R_2$.} The RRMSE results for this geometry-transfer test are summarized in Table~\ref{tab:rrmse_all}. The very low RRMSE values are remarkable. For all the fields ($||\mathbf{v}||$, $c_1$, $c_2$, and $||\bm \tau||$) the RRMSE values remain below 1 \%.


Figure \ref{fig:c1_c2_map_crossslot} shows the $c_1$-$c_2$ map for cross-slot simulations at Wi=0.5 (red crosses) and Wi=1.0 (orange triangles). The training points gathered from a flow around a cylinder Giesekus-FV simulation at Wi=5.0 are shown in blue and allow us to determine {whether} the $c_1$-$c_2$ space explored by the cross-slot geometry simulations falls inside the training hull. For Wi=0.5, the space explored is compatible with the training and as a result, the agreement between FV-Giesekus (red) and FV-GINNGER (green) is excellent. To summarize, in our method the training region $c_1$-$c_2$ space determines what simulations can be accomplished regardless of the type of geometry. In this sense, the GINNGER methodology is geometry-agnostic, but it is limited in accuracy to the training space explored in the $c_1$-$c_2$ space which is completely general (i.e., non-geometry-specific). 
{As a matter of fact}, at {Wi=1.0} this test exposes the limitation of the transfer strategy: the
{$\Wi=1.0$} cross-slot case does not converge numerically since the simulation probes $c_1$-$c_2$ states that are well outside the manifold explored by the {$\Wi=5.0$} flow around a cylinder training. In particular, the methodology is more sensitive to larger maximum values for $c_1$ (up to $\sim20$) that are encountered during cross-slot flow at Wi=1.0. In the simulation used for training -- flow around a cylinder at $\Wi=5.0$ -- $c_1$ reaches a maximum value around $\sim 18.5$. As in the case for flow around a cylinder, requesting predictions from the GINNGER model at higher $c_1$ values can result in an overly demanding extrapolation {that poses notorious difficulties in PINNs \cite{Bonfanti2024a}}. Note that, even for Wi=0.5, a few points at moderate $c_1$ and $c_2$ values are significantly outside the training domain in Figure \ref{fig:c1_c2_map_crossslot} due to the small numerical oscillations in the cross-slot corners. These are reasonably well captured by the FV-GINNGER model as a result of the decent representation of the constitutive model given by $\sigma^\NN_\alpha$ and $R_{2,\alpha}^\NN$ in these intermediate regions even when outside the training domain (see Figures \ref{fig:training_results_stage1_wi5p0} and \ref{fig:training_results_stage2_wi5p0}).

\section{Discussion}
\label{sec:discussion}

In this article, we have shown that an \emph{a priori} unknown constitutive equation for a single relaxation time viscoelastic fluid can be learned using the GENERIC-Informed Neural Network for GEnerative Rheology (GINNGER). With this new procedure, we provide a viable route to learning constitutive behavior directly from polymer conformation, stress and kinematic fields ($||\mathbf{v}||$, $c_1$, $c_2$, and $||\bm \tau||$) in complex flow situations where the analytical form of the constitutive law is unknown.
Rather than identifying constitutive equations directly, GINNGER reformulates constitutive discovery as the learning of the thermodynamic generators represented by the neural-network approximations $\bm{\sigma}^{\rm NN}$ and $\bm{R}_2^{\rm NN}$, from which admissible constitutive equations naturally emerge.
The methodology relies on the universal learning of both the entropy and the friction matrix functionals in the $\bm c$-space without prescribing any closed-form constitutive law.
Furthermore, this learning is performed in the intrinsic spectral state space of the conformation tensor rather than in flow-dependent physical coordinates, allowing constitutive behavior to be characterized in terms of the underlying microstructural state.
In this sense, the GINNGER approach deviates from SINDY-type strategies of pre-selected model identification \cite{discovering_brunton_2016}.

In this novel GINNGER methodology the three key assumptions are: (i) the effect of microstructure on the flow dynamics can be described by a 2nd rank tensorial closure given by the conformation tensor $\bm c$ and the friction matrix $\bm R_2$; (ii) the flow-deformation is considered affine; and (iii) that the fluid models considered here are purely entropic. Regarding assumption (ii), we have shown that non-affine deformation effects could be learned from data and incorporated as a non-entropic production term in a modified mixed derivative \cite{Nieto2023}. To the best of our knowledge, this is the first work presenting a two-stage GENERIC-guided learning strategy for the GINNGER framework and related CFD-RheoTool implementation that is general and preserves the thermodynamic structure by design. This aspect is crucial for several reasons. First, only thermodynamically consistent models survive the data-driven searching criteria, which is critical to avoid unphysical modelling issues \cite{Schieber2021}. Secondly, optimization (i.e. NN loss minimizations) is done on subspaces of the whole parameter space, which is numerically convenient.
More fundamentally, the GENERIC decomposition shifts the learning problem from the identification of high-dimensional constitutive operators to the identification of a small number of thermodynamic generators, thereby substantially reducing the complexity of the search space.
Third, interpretability of the neural network is improved, since we can monitor convergence on specific thermodynamic blocks with sound physical meaning, rather than arbitrary multi-dimensional operators whose meaning is numerically hidden.

The learning procedure -- constrained by the GENERIC structure and therefore physics-informed in a thermodynamic sense -- is based on three scalar-regression neural networks. In the first stage, one network parametrizes the entropy potential (and thus the eigenvalues of $\bm \sigma^\NN$). In the second stage, two additional networks parametrize the eigenvalues of $\bm R_2^\NN$.
Together, $\bm{\sigma}^{\rm NN}$ and $\bm{R}_2^{\rm NN}$ constitute neural representations of the thermodynamic generators defining the constitutive model, rather than neural approximations of the constitutive equation itself.
By working with the eigenvalues of the conformation tensor $\bm c$, the thermodynamic conjugate $\bm \sigma$ and the friction matrix $\bm R_2$, all regressions are scalar and no direct tensor regression is required, reducing the computational cost which scales with the problem dimensionality.

More importantly, this spectral representation provides the intrinsic thermodynamic state space in which the thermodynamic generators are naturally defined. In this formulation, the flow kinematics merely transports the polymer microstructure through the spectral state space, whereas the constitutive information depends only on the instantaneous microstructural state. Consequently, shear, extensional, and mixed flows no longer constitute separate learning problems but rather different trajectories through the same conformation state space, allowing heterogeneous datasets to be combined to identify a single constitutive model.
Working in the $\bm{c}$-eigenspace is also useful to enforce material objectivity of the NN approximation since, for example, $s_c^{\NN}(c_1,c_2)$ does not manifestly depend on the reference frame.

To validate and test the limits of our GINNGER methodology, we use the Giesekus model as example. This model features both a non-trivial entropy and non-linear friction matrix that result in the rich rheological behavior typical of melts and concentrated solutions. Embedded cell-by-cell in RheoTool/OpenFOAM, FV-GINNGER is able to recover the reference Giesekus solution for the flow around a cylinder at Weissenberg numbers below or slightly above that of the training. For the error assessment we combine global RRMSE and damped relative-error fields to separate genuinely relevant deviations from artifacts near vanishing stresses.

The CFD FV-GINNGER model, trained in flow around a cylinder, remains accurate when applied to a cross-slot geometry without retraining, provided that the explored $c_1$--$c_2$ space is compatible with the trained NN.
This observation highlights one of the key consequences of the proposed spectral formulation: the predictive capability of the learned constitutive model is governed by the coverage of the spectral thermodynamic state space rather than by the particular flow geometry or deformation protocol used during training. As a result, transferring a constitutive model between different flow types should not be regarded as extrapolation whenever they explore the same region of this state space, but rather as interpolation between previously learned thermodynamic states.
This viewpoint, originally introduced in \cite{Nieto2025}, provides also a natural interpretation of the recently identified kinematic and rheological equivalence between steady shearing and planar extensional flows \cite{king2026}.

This geometry-agnostic character of the methodology represents a significant advancement over current machine learning methods applied to rheological constitutive modeling. At the same time, our tests expose two limitations to be addressed in future work. First, the current neural-network architecture exhibits instabilities in highly stretched regions, motivating the need for more expressive architectures or additional regularization methods. Second, the cross-slot case at larger Wi fails to converge because parts of the solution probe $(c_1,c_2)$ lie outside the manifold explored by the training data.
In this framework, extrapolation therefore acquires a precise physical meaning: it occurs only when the microstructure evolves into regions of the spectral thermodynamic state space that were not represented during training, independently of the kinematics that generated them.
These results highlight the importance of coverage in conformation space and the choice of the training geometry and Wi, which can be improved by combining data from multiple complementary experiments \cite{Sankaran2026}.

\section{Conclusions}
\label{sec:conclusions}
We introduce the GENERIC-Informed Neural Network for GEnerative Rheology (GINNGER), a framework that reformulates constitutive discovery as the learning of thermodynamic generators from which constitutive equations emerge, while ensuring thermodynamic consistency by construction. By embedding the GENERIC structure directly into the learning process, the approach reduces model complexity, improves interpretability, and avoids unphysical solutions. Furthermore, by performing learning in the intrinsic spectral state space of the conformation tensor rather than in flow-dependent physical coordinates, GINNGER naturally separates the thermodynamic state of the material from the external kinematics. The resulting models generalize across flow conditions and geometries, provided the relevant spectral microstructural state space is adequately sampled.

This spectral formulation also provides a new physical interpretation of interpolation and extrapolation in constitutive learning. Different deformation protocols correspond simply to different trajectories through the same microstructural state space, whereas genuine extrapolation occurs only when previously unseen thermodynamic states are explored. Consequently, the predictive capability of the learned constitutive model is governed by the coverage of the explored microstructural state space rather than by the particular flow geometry or kinematics used to generate the training data.

These results establish a generative and spectral paradigm for data-driven constitutive modeling of complex fluids, in which thermodynamic principles guide both the learning process and the definition of the admissible hypothesis space, going beyond traditional constitutive model selection. More broadly, GINNGER suggests that constitutive behavior is most naturally organized by the intrinsic thermodynamic state of the material rather than by the external deformation protocol. This framework opens the door to learning physically admissible constitutive laws directly from simulations or experiments, with broad implications for scientific machine learning, computational mechanics, and nonequilibrium thermodynamics.

\begin{acknowledgments}
M.E. and I.G.B. acknowledge the funding provided by IKUR–HPC\&AI – (HPCAI10: MOLD-POLY) and the IKUR Strategy funded by Basque Government and the European Union NextGenerationEU/PRTR. The research is also partially funded by the Spanish State Research Agency through BCAM Severo Ochoa excellence accreditation CEX2021-0011 42-S/MICIN/AEI/10.13039/501100011033, and through the project PID2024-158994OB-C42 (‘Multiscale Modeling of Friction, Lubrication, and Viscoelasticity in Particle Suspensions’ and acronym ‘MMFLVPS’) funded by MICIU/AEI/10.13039/501100011033. I.G.B. also acknowledges the support by the grant from the Department of Science, University and Innovation from the Basque Government to the MATHMODE Group (IT1866-26). The authors declare no competing interest.
\end{acknowledgments}

\section*{Author Contributions}
D.N.S. and M.E. designed research; I.G.d.B. developed software; I.G.d.B. and D.N.S. performed research; and I.G.d.B., D.N.S., and 
M.E. analyzed data and wrote the paper.

\FloatBarrier

\appendix

\section{GENERIC formulation of constitutive models}
\label{sec:GENERIC_formulation}
\subsection{Oldroyd-B model closure in \textsc{GENERIC}}
For a dilute suspension of Hookean dumbbells, Sarti and Marrucci have derived an analytical expression for $s_c({\bm c})$ \cite{sarti1973thermomechanics}.
\begin{equation}
      s_c({\bm c})
 =  \frac{k_B n_p}{2}  \left[ {\rm tr}({\bm 1}-{\bm c})  + \ln(\det {\bm c}) \right]  \label{eq:entropy_OB}
\end{equation}
where $k_B$ is the Boltzmann constant, $n_p$ is the constant concentration of suspended polymers and ${\bm 1}$ is the identity matrix. This expression for the entropy can be derived bottom-up from microscopic systems by means of statistical mechanics \cite{vazquez2009b, Pavelka2018}. Note that the dependence of $s_c$ on invariants of $\bm c$, like ${\rm tr}({\bm c})$ and $ {\rm det}({\bm c})$, is crucial for the``material objectivity" of the entropy function, which arises as a consequence of the \textsc{GENERIC} framework {(Appendix \ref{app:app2})}.  If we differentiate the entropy expression in Eq. (\ref{eq:entropy_OB}), we get the thermodynamic conjugated variable to $\bm c$, defined as
\begin{equation}
       \frac{\bm \sigma}{T} = \frac{\partial s_c({\bm c})}{\partial {\bm c}}
 =  \frac{k_B n_p}{2}  \left[ {\bm c}^{-1}  - {\bm 1} \right]  \label{eq:sigma_OB}
\end{equation}
Inserting Eq. (\ref{eq:sigma_OB}) into Eq. (\ref{eq:generic_stress}) leads to the classical {Oldroyd-B stress relation}
\begin{equation}
 {\bm \tau} ^{{\rm OB}}({\bm c})
 =  {k_B T n_p} \left[ {\bm 1}- {\bm c}  \right] = \frac{\eta_p}{\lambda}  \left[ {\bm 1}- {\bm c}  \right]   \label{eq:stress_OB}
\end{equation}
where we have defined the {polymeric viscosity} $\eta_p=k_B T n_p \lambda$ and $\lambda$ is the polymer's longest relaxation time.\\

Similarly, for the evolution equation for ${\bm c}$ we use Eq. (\ref{eq:generic_c2}) to get
\begin{equation}
\ucd{\bm c} 
 = {{\bm R}_2}  :  \frac{k_B n_p}{2}  \left[ {\bm c}^{-1}  - {\bm 1} \right]  \label{evol_OB1}
\end{equation}

Finally, under the assumption of non-hydrodynamically interacting polymers (by assuming a purely white noise Wiener process in the Langevin dynamics for each end-to-end vector ${\bf q}$), it is possible to infer the stochastic noise property on the tensor ${\bf c}$ and therefore the friction matrix ${\bm R}_2$ directly from it as a 4$^{th}$-rank tensor \cite{vazquez2009},  
%
%\footnote{The derivation can be obtained directly in the case of a suspension on dilute non-interacting polymers, by assuming a Langevin dynamics for the effective end-to-end vector ${\bm q}$ of the polymer chain (i.e. we consider on the first slowest Rouse mode in a dumbbell approximation of the chain), and assume that diffusional property are thus related only to the thermal bath. It is possible to infer analytically the diffusional properties for the conformation tensor. Details provided in Sec...}
%
\begin{equation}
        {{R}_2}^{ijlk}
 =  \frac{1}{2 n_p k_B \lambda} \left[  {c}_{il}  {\delta}_{jk} +  {c}_{lj}  {\delta}_{ik} +  { c}_{ik}  {\delta}_{jl} +  {c}_{jk}  { \delta}_{il} \right] 
 \label{Mdil}
\end{equation}
which, once contracted with Eq.(\ref{evol_OB1}), leads to
\begin{equation}
       \ucd{\bm c} 
 = \ \frac{1}{\lambda} ( {\bm 1} - {\bm c})   \label{evol_OB2}
\end{equation}
Alternatively, it has been shown that the 4$^{th}$-rank tensor ${{\bm R}_2}^{ijlk}$ can be replaced by 2$^{nd}$-rank tensor 
${{\bm R}_2}^{ij}=\sum_l{{\bm R}_2}^{ijll}$ \cite{Ottinger_book}. In that case ${{\bm R}_2}^{ij}= \frac{2}{n_p k_B \lambda}  {\bm c}$ and the final { Oldroyd-B evolution equation} again reads

\begin{equation}
     \ucd{\bm c} 
 =    \frac{2}{n_p k_B \lambda} {\bm c} \cdot    \frac{\partial s_c({\bm c})}{\partial {\bm c}} =  \frac{1}{\lambda} ( {\bm 1} - {\bm c})   \label{evol_OB3}
\end{equation}

\subsection{Giesekus model closure in \textsc{GENERIC}}
The Giesekus model assumes the same entropy function for the Oldroyd-B model \cite{Giesekus1982,GIESEKUS1983,Schieber2021}. Considering an arbitrary choice for ${\bm R}_2$ is equivalent to shifting physically from the {\em dilute} (non-hydrodynamically interacting polymers) towards more {\em concentrated} or dense systems, where the dynamics can be strongly non-linear and anisotropic. To that end, Giesekus proposed a purely phenomenological friction matrix form \cite{GIESEKUS1983,Giesekus1982}
\begin{equation}
   {{\bm R}_2}
 =    \frac{2}{n_p k_B \lambda} \left[ (1-\alpha) {\bm c} + \alpha {\bm c}^2  \right]     \label{R2_G}
\end{equation}
which leads to the {Giesekus evolution equation} for the conformation tensor

\begin{equation}
     \ucd{\bm c} 
 =    \frac{2}{n_p k_B \lambda}   \left[ (1-\alpha) {\bm c} + \alpha {\bm c}^2  \right]     \cdot    \frac{\partial s_c({\bm c})}{\partial {\bm c}}   \label{evol_GIESEKUS}
\end{equation}
where $\alpha$ is a phenomenological parameter which, for thermodynamic consistency,  needs to be $0\le \alpha \le 1$. Since the \textsc{GENERIC} stress definition in Eq.(\ref{eq:generic_stress}) depends only on the entropy, which is the same as in the Oldroyd-B model, the Giesekus stress relation is exactly the same as in Eq. (\ref{eq:stress_OB}),
  \begin{equation}
     {\bm \tau} ^{{\rm Giesekus}}({\bm c})
 =  {k_B T n_p} \left[ {\bm 1}- {\bm c}  \right] = \frac{\eta_p}{\lambda}  \left[ {\bm 1}- {\bm c}  \right]   \label{eq:stress_GIESEKUS}
\end{equation}

\subsection{GENERIC microstructure-based models in the {\bf c}-eigenbasis}
Let us diagonalize the conformation tensor in the following way
  \begin{equation}
    {\bm c} = \sum_\alpha c_\alpha {\bm u}_\alpha   {\bm u}_\alpha  \label{eq:diag}
\end{equation}
where $c_\alpha$  and ${\bm u}_\alpha$ are eigenvalues and eigenvectors and $\alpha=1,.., D$. The following properties hold
 \begin{equation}
    {\bm c} {\bm u}_\alpha  = c_\alpha {\bm u}_\alpha   
\end{equation}
 and the orthonormalization of the eigen-basis:
 \begin{equation}
    {\bm u}_\alpha \cdot   {\bm u}_\beta = \delta_{\alpha\beta}   \label{ort}
\end{equation}
and
\begin{equation}
   \frac{d}{dt} \left(  {\bm u}_\alpha \cdot   {\bm u}_\beta   \right )= 0   \implies \dot{\bm u}_\alpha {\bm u}_\beta = -   {\bm u}_\alpha \dot {\bm u}_\beta  \label{norm}
\end{equation}
\newline
Since ${\bm \sigma}$ is a tensorial function of ${\bm c}$, it diagonalizes on the same eigenbasis and therefore we can define its eigenvalues as
 \begin{equation}
    {\bm \sigma} {\bm u}_\alpha  = \sigma_\alpha {\bm u}_\alpha   
\end{equation}

Taking the time derivative of Eq.(\ref{eq:diag}), where $\dot{(..)}=d/dt (..)$ denotes the Lagrangian derivative, leads to
  \begin{equation}
    \dot{\bm c} = \sum_\gamma \dot{c}_\gamma {\bm u}_\gamma   {\bm u}_\gamma  + c_\gamma  \left(  \dot{\bm u}_\gamma  {\bm u}_\gamma  +  {\bm u}_\gamma  \dot {\bm u}_\gamma \right)
\end{equation}
 Now, we right/left multiply it by the eigenvectors, (i.e. ${\bm u}_\alpha  \cdot (..) \cdot {\bm u}_\beta$), to get
   \begin{equation}
  {\bm u}_\alpha  \cdot  \dot{\bm c} \cdot {\bm u}_\beta = \dot{c}_\alpha  \delta_{\alpha\beta} + \left(c_\alpha - c_\beta  \right) \dot {\bm u}_\alpha \cdot {\bm u}_\beta \label{diag1} 
\end{equation}
Note that in the most general case, when we project the general evolution equation (\ref{eq:generic_c2}) into the eigenvector basis, we have 
\begin{equation}
  {\bm u}_\alpha  \cdot   \dot{{\bm c} } \cdot {\bm u}_\beta=  c_\beta \kappa_{\alpha\beta} + c_\alpha \kappa_{\beta\alpha}
 +   R_2^{\alpha\beta}  \sigma_\beta 
\end{equation}
where
\begin{equation}
 R_2^{\alpha\beta}= {\bm u}_\alpha \cdot\!{\bm R}_2 \cdot {\bm u}_\beta\\
 \end{equation}
However, since ${\bm R}_2$ diagonalizes in the same basis as $\bm c$, we have 
\begin{equation}
  {\bm u}_\alpha  \cdot   \dot{{\bm c} } \cdot {\bm u}_\beta=  c_\beta \kappa_{\alpha\beta} + c_\alpha \kappa_{\beta\alpha}
 +   R_2^{\alpha}  \sigma_\beta \delta_{\alpha\beta} 
\end{equation}
Here $R_2^{\alpha}$ are the eigenvalues of ${\bm R}_2$.
In general a tensor ${{\bm R}_2}({\bm c})$ will diagonalize in the same eigenbasis as ${\bm c}$ when it preserves the eigenspace decomposition of ${\bm c}$. This occurs when ${\bm R}_2$ commutes with ${\bm c}$ (i.e. the following commutator condition holds: $[{{\bm R}_2}({\bm c}),{\bm c}]=0$).
 
 For the purpose of validating our methodology, we will use the Giesekus model given by Eqs (\ref{eq:entropy_OB}) and (\ref{R2_G}) as an example. In the case of Giesekus, we have that ${\bm R}_2$ in Eq.(\ref{R2_G}) indeed does diagonalize on the ${\bm c}$-eigenbasis since it is an analytical function 
of $\bm c$. As a result,  the { evolution equation for the eigenvalues in the Giesekus  model} reads
\begin{equation}
\dot{c}_\alpha  = 2 c_\alpha \kappa_{\alpha\alpha}
 +    \underbrace{ \frac{2}{n_p k_B \lambda}    \left[ (1-\alpha) {\bm c_\alpha} + \alpha {\bm c_\alpha}^2  \right]}_{R_2^\alpha}  \sigma_\alpha \label{eq:eigenvalue_G}
\end{equation}
This expression can be further simplified using the eigenvalues of $\bm \sigma$ in Eq. (\ref{eq:sigma_OB})
\begin{equation}
\dot{c}_\alpha  =   \frac{1}{\lambda} \left( {1} - {c}_\alpha \right) \left[ (1-\alpha)  + \alpha {c}_\alpha  \right] \label{eq:eigenvalue_G2}
\end{equation}
Note that, in the case of Oldroyd-B the friction matrix ${\bm R}_2\propto{\bm c}$ and it trivially diagonalizes on the same basis as $\bm c$. The Oldroyd-B model presents an additional simplification since the dynamics of each eigenvalue of $\bm \sigma$ depend only on the corresponding eigenvalues of $\bm c$ (i.e., $\sigma_\alpha=\sigma_\alpha(c_\alpha)$). The key difference for more complex models, like FENE-P, is that the dynamics for all the eigenvalues are mixed {through the trace of $\bm{c}$} \cite{Nieto2025}.

\section{Flow around a cylinder validation at lower Wi}\label{app:app1}
\begin{figure}[hbt!]
  \centering
  \begin{center}
  \begin{tabular}{cc}
      \includegraphics[width=0.4\linewidth]{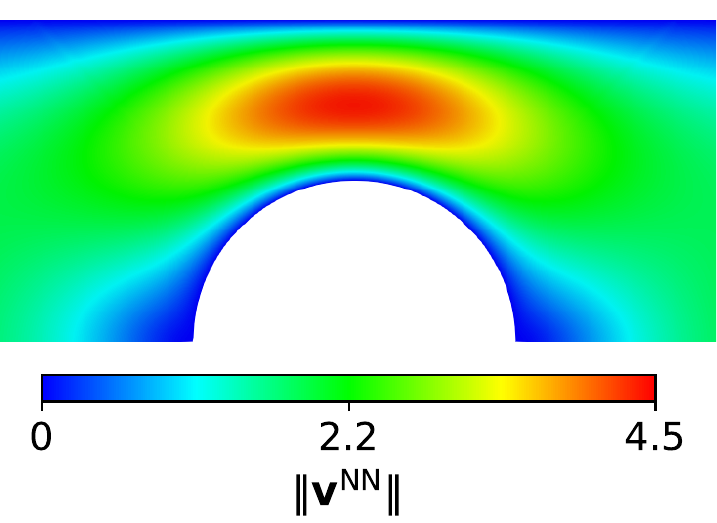} &  \includegraphics[width=0.4\linewidth]{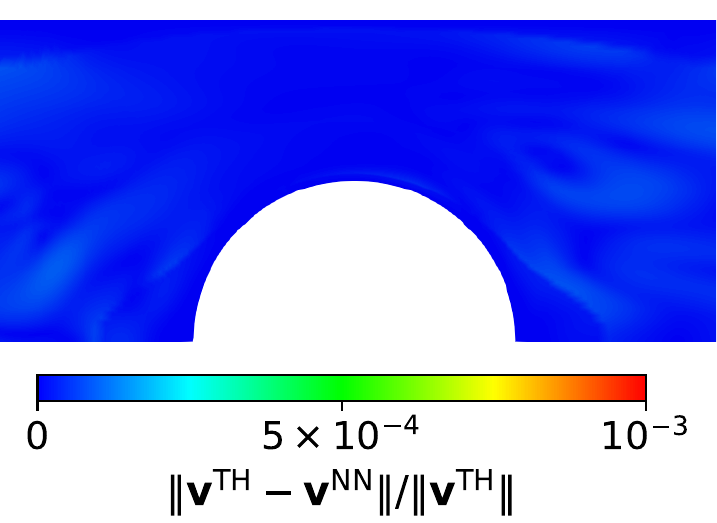}\\
       \includegraphics[width=0.4\linewidth]{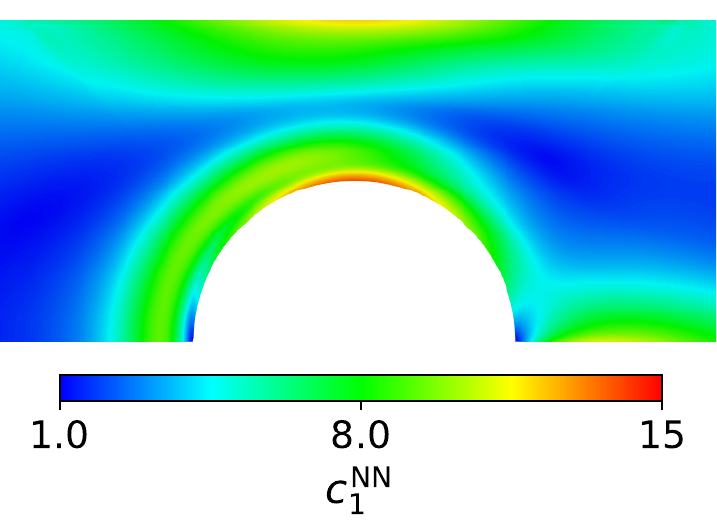} & \includegraphics[width=0.4\linewidth]{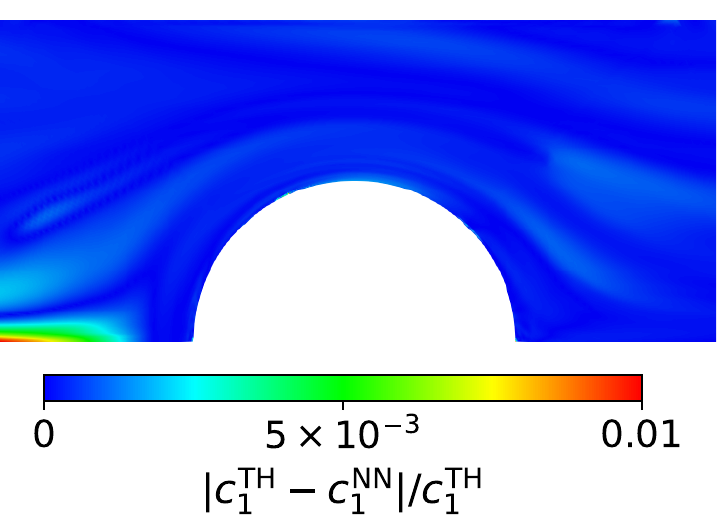}\\
       \includegraphics[width=0.4\linewidth]{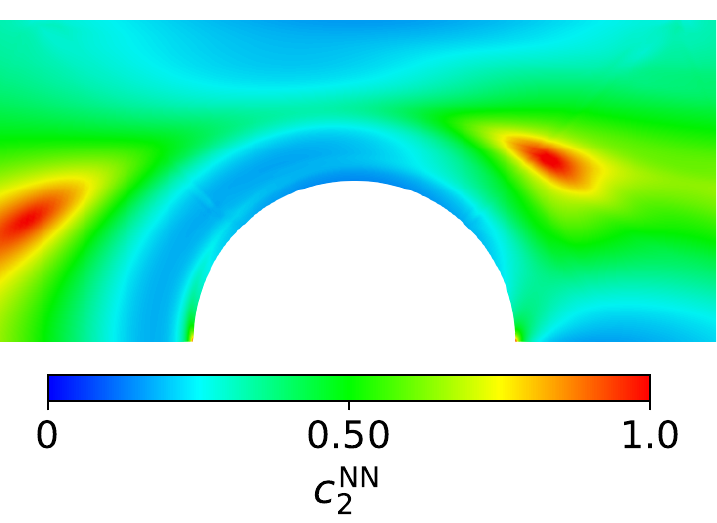} & 
       \includegraphics[width=0.4\linewidth]{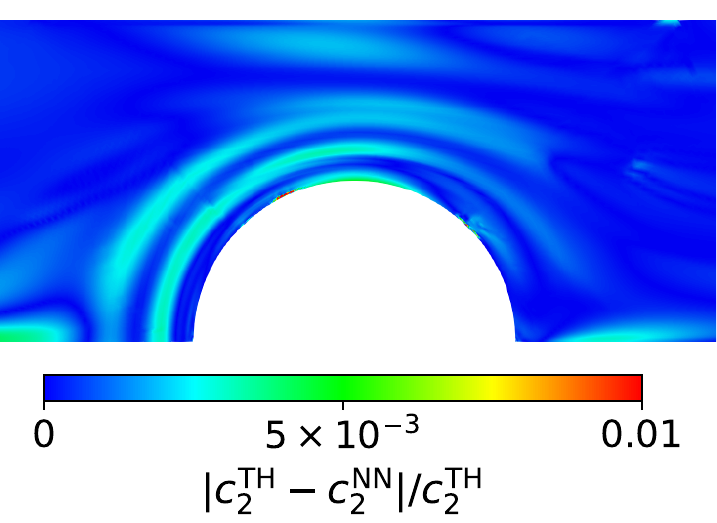}\\
       \includegraphics[width=0.4\linewidth]{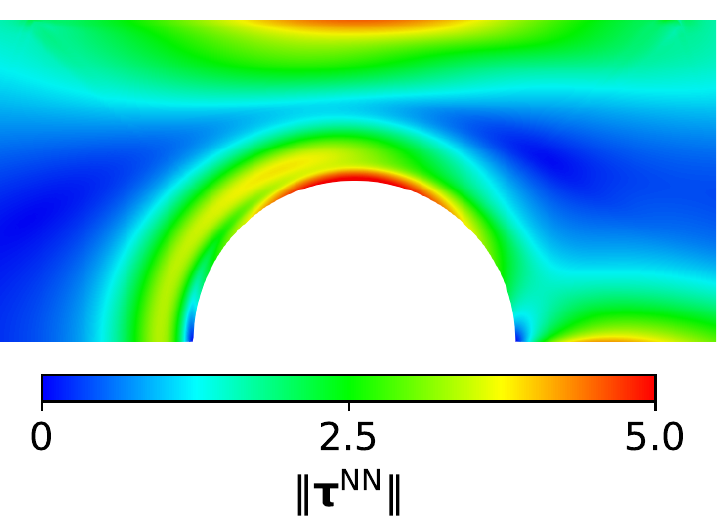} & \includegraphics[width=0.4\linewidth]{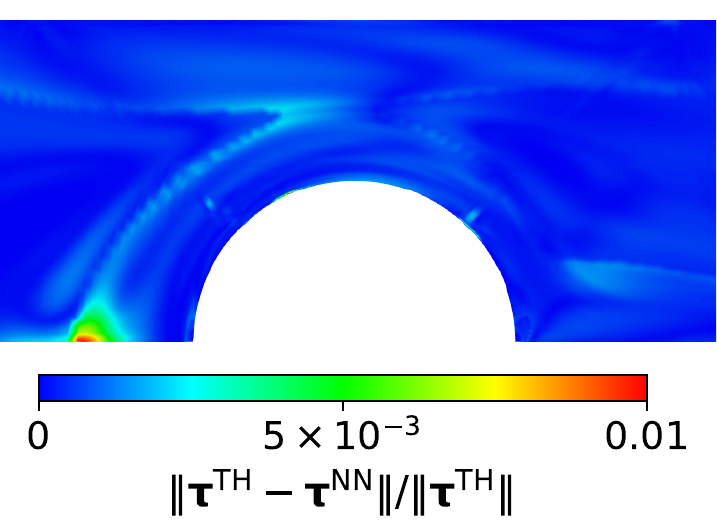}
  \end{tabular}
    \end{center}
\caption{GINNGER-FV simulation results for $U$, $c_1$, $c_2$, and $\tau$ fields (left) and their relative-error fields (right) in flow around a cylinder at $\Wi=3$.}
  \label{fig:wi3p0_fields}
\end{figure}

\begin{figure}[hbt!]
  \centering
  \begin{tabular}{cc}
    \includegraphics[width=0.8\linewidth]{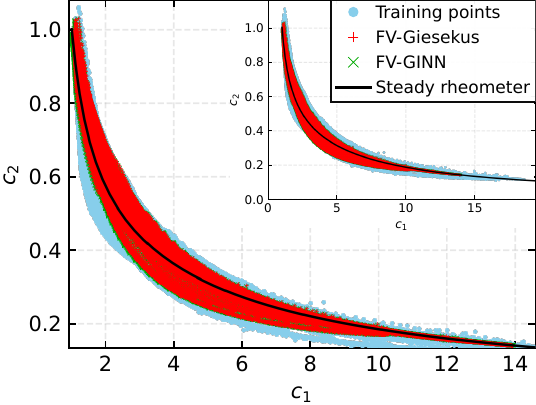}
  \end{tabular}
  \caption{{$c_1$-$c_2$ maps for the flow-around-cylinder validation at $\Wi=3.0$.}}
  \label{fig:c1_c2_map_cylinder_Wi3.0}
\end{figure}

In the main text of the manuscript, we have presented results showing that the GINNGER model resulting from training with data from a complex flow simulation at a given flow rate (or Wi number) can generally be applied in simulations of the same flow geometry at lower rates. An additional example is given in Figures \ref{fig:wi3p0_fields} and \ref{fig:c1_c2_map_cylinder_Wi3.0}. The results for the fields $||\mathbf{v}||$, $c_1$, $c_2$, and $||\bm \tau||$ and their relative errors in Figure \ref{fig:wi3p0_fields} show excellent agreement with simulations using the Giesekus model in Rheotool. The good performance of the model at lower Wi is a direct result of the region explored in the $c_1$-$c_2$ space being inside the domain used for training (see Figure \ref{fig:c1_c2_map_cylinder_Wi3.0}). 

\newpage
\newpage

{
\section{Material objectivity of the eigen-basis spectral NN representation}\label{app:app2}

Let ${\bm c}$ be a symmetric $3\times 3$ second-order tensor. As a consequence of the Cayley-Hamilton theorem 
we have that any tensorial function ${\bm R}({\bm c})$ can be expressed as
\[
{\bm R}({\bm c}) = \beta_0 {\bm 1} + \beta_1 {\bm c} + \beta_2 {\bm c}^2,
\]
where the coefficients $\beta_i$ are functions of the invariants of ${\bm c}$, that is  $I_1, I_2, I_3$ which in 3D space read
\[
I_1 = \mathrm{tr}({\bm c}), \quad
I_2 = \tfrac12 \left[(\mathrm{tr} {\bm c})^2 - \mathrm{tr}({\bm c}^2)\right], \quad
I_3 = \det({\bm c})
\]
This ensures {\em material objectivity}, since invariants do not change under any proper orthogonal rotation ${\bm Q}$, leading to
\[
{\bm c} \to {\bm c}^* = {\bm Q} {\bm c} {\bm Q}^{\mathrm{T}} 
\quad \Rightarrow \quad 
{\bm R}({\bm c}) \to {\bm R}^* = {\bm Q} {\bm R}({\bm c}) {\bm Q}^{\mathrm{T}}.
\]
Now, using the Cayley-Hamilton representation as a \emph{tensorial basis}, a neural network can learn a tensor function of ${\bm c}$:
\[
{\bm R}^{\mathrm{NN}}({\bm c}) = \beta^{\mathrm{NN}}_0(I_1,I_2,I_3) {\bm 1} + \beta^{\mathrm{NN}}_1(I_1,I_2,I_3) {\bm c} + \beta^{\mathrm{NN}}_2(I_1,I_2,I_3) {\bm c}^2
\]
where the scalar functions $\beta^{\mathrm{NN}}_i$ are outputs of a neural network taking the invariants $(I_1,I_2,I_3)$ as input data.  
By construction, the {\em tensorial basis neural network} (TBNN) representation ${\bm R}^{\mathrm{NN}}$ is thus forced to be objective \cite{scientific_lennon_2023}. \\
As an alternative, in this work we have considered a {\em spectral eigen-basis neural network} representation. Again, let us assume that,  ${\bm c}$ is symmetric, as already discussed, it can be diagonalized as
\[
{\bm c}\, {\bm u}_\alpha= c_\alpha\, {\bm u}_\alpha, \quad \alpha=1,2,3,
\]
where $\{{\bm u}_1, {\bm u}_2, {\bm u}_3\}$ form an orthonormal basis and $c_1, c_2, c_3$ are the eigenvalues.\\
The spectral neural network representation of a general tensor function ${\bm R}({\bm c})$ reads therefore
\[
{\bm R}^{\mathrm{NN}}({\bm c}) = \sum_{\alpha=1}^{3} R^{\mathrm{NN}}_\alpha(c_1,c_2,c_3)\, {\bm u}_\alpha  {\bm u}_\alpha, 
\]
where we have used the fact that ${\bm R}({\bm c})$ share the same eigenbasis of ${\bm c}$ and 
each $R^{\mathrm{NN}}_\alpha$ is a scalar function of the eigenvalues.  This NN representation trivially satisfies objectivity in the non-degenerate case (in 2D it would be only at equilibrium for $c_1=c_2=1$). In fact:
\[
{\bm c} \to {\bm c}^* = {\bm Q} {\bm c} {\bm Q}^{\mathrm{T}}, \quad
{\bm u}_i \to {\bm u}_i^* = {\bm Q} {\bm u}_i, \quad
c_i^* = c_i,
\]
such that
\[
\begin{aligned}
{\bm R}^{\mathrm{NN*}}({\bm c}^*)
&= \sum_i R_i(c_1,c_2,c_3)\, {\bm u}_i^* {\bm u}_i^* \\
&= {\bm Q} \left( \sum_i R_i(c_1,c_2,c_3)\, {\bm u}_i {\bm u}_i \right) {\bm Q}^{\mathrm{T}} \\
&= {\bm Q}\, {\bm R}^{\mathrm{NN}}({\bm c})\, {\bm Q}^{\mathrm{T}}.
\end{aligned}
\]
This shows also that the two representations are equivalent by simply rewriting the invariants in terms of $c_\alpha$. That is:
\[
I_1 = c_1 + c_2 + c_3, \quad
I_2 = c_1 c_2 + c_2 c_3 + c_3 c_1, \quad
I_3 = c_1 c_2 c_3.
\]
 Similar arguments apply to a scalar function  (such as the polymeric entropy). In particular, let $s({\bm c})$ be a scalar function of the tensor ${\bm c}$.  Objectivity requires 
\[
s({\bm c}^*) = s({\bm c}) \quad \text{for all } {\bm Q} \in SO(3).
\]
By using the eigen-basis spectral NN decomposition adopted in this work  we have
\[
s^{\mathrm{NN}}({\bm c}) = s^{\mathrm{NN}} (c_1, c_2, c_3)
\]
and, since the eigenvalues are invariant under rotations, it necessarily implies that $s^{\mathrm{NN}} (c_1, c_2, c_3) $ %(and its conjugate variable ${\bm \sigma}^{\mathrm{NN}}$) 
is by construction invariant too in the non-degenerate case. Conversely, in a non-diagonalized ${\bm c}$-space, in order to have rotational invariance, the entropy should be forced to depend on ${\bm c}$-invariants $I_1,I_2,I_3$ which is consistent with analytical expressions for the polymeric entropy in Eq.(\ref{eq:entropy_OB}), for example. 

}

\bibliography{PINNGiesekus}

\end{document}